\documentclass[a4paper,aps,twocolumn,showpacs,prb,superscriptaddress
]{revtex4-1}

\usepackage[utf8]{inputenc}
\usepackage[T1]{fontenc}
\usepackage{microtype}
\usepackage{mathbbol}
\usepackage{amssymb}
\usepackage[usenames,dvipsnames,svgnames,table]{xcolor}
\usepackage[colorlinks,linkcolor=blue,citecolor=blue,urlcolor=blue]{hyperref}

\usepackage{graphicx}
\usepackage{amssymb}%,bbm}
\usepackage{amsmath,amsfonts,latexsym}
\usepackage{array,tabularx}
\usepackage{dcolumn}               % Align table columns on decimal point
\usepackage{fancyhdr}
\usepackage{hyperref}
\usepackage{comment}
\usepackage{placeins}

\newcommand{\braket}[1]{\ensuremath{\langle{#1}\rangle}}
\newcommand{\bra}[1]{\langle #1 |}
\newcommand{\ket}[1]{| #1 \rangle}

\def\equationautorefname~#1\null{Eq.~(#1)\null}

\begin{document}
	
	\title{Tensor Network simulation of polaron-polaritons in organic microcavities}
	
	\author{Javier del Pino}
	\email{francisco.delpino@uam.es}
	\affiliation{Departamento de Física Teórica de la Materia Condensada and Condensed Matter Physics Center (IFIMAC), Universidad Autónoma de Madrid, E-28049 Madrid, Spain}
	\author{Florian A. Y. N. Schröder}
	\affiliation{Cavendish Laboratory, University of Cambridge, J. J. Thomson Avenue, Cambridge, CB3 0HE, UK}
	\author{Alex W. Chin}
	\affiliation{Institut des NanoSciences de Paris, Sorbonne Université, 4 place Jussieu, boîte courrier 840, 75252, PARIS Cedex 05}
	\affiliation{Cavendish Laboratory, University of Cambridge, J. J. Thomson Avenue, Cambridge, CB3 0HE, UK}
	\author{Johannes Feist}
	\affiliation{Departamento de Física Teórica de la Materia Condensada and Condensed Matter Physics Center (IFIMAC), Universidad Autónoma de Madrid, E-28049 Madrid, Spain}
	\author{Francisco J. Garcia-Vidal}
	\affiliation{Departamento de Física Teórica de la Materia Condensada and Condensed Matter Physics Center (IFIMAC), Universidad Autónoma de Madrid, E-28049 Madrid, Spain}
	\affiliation{Donostia International Physics Center (DIPC), E-20018
		Donostia/San Sebastián, Spain}
	
	\begin{abstract}
		In the regime of strong coupling between molecular excitons and confined optical modes,
		the intra-molecular degrees of freedom are profoundly affected, leading to a reduced
		vibrational dressing of polaritons compared to bare electronically excited states.
		However, existing models only describe a single vibrational mode in each molecule, while
		actual molecules possess a large number of vibrational degrees of freedom and
		additionally interact with a continuous bath of phononic modes in the host medium in
		typical experiments. In this work, we investigate a small ensemble of molecules with an
		arbitrary number of vibrational degrees of freedom under strong coupling to a
		microcavity mode. We demonstrate that reduced vibrational dressing is still present in
		this case, and show that the influence of the phononic environment on most electronic and
		photonic observables in the lowest excited state can be predicted from just two collective parameters of the
		vibrational modes. Besides, we explore vibrational features that can be addressed
		exclusively by our extended model and could be experimentally tested. Our findings
		indicate that vibronic coupling is more efficiently suppressed for environments characterised
		by low-frequency (sub-Ohmic) modes. 
	\end{abstract}
	
	\maketitle
	
	\section{Introduction}
	
		When an ensemble of quantum emitters interacts with a confined electromagnetic field (EM)
	mode, the system can enter into the strong coupling regime, resulting in the formation of
	light-matter quasiparticles known as polaritons, which inherit properties from each
	constituent. In particular, organic molecules present favourable features to achieve
	large couplings to optical modes at room temperature, due to their tightly-bound Frenkel
	excitons~\cite{Davydov1971,Agranovich2008} with large dipole moments. These properties
	offer an optimal experimental platform for polariton lasing
	\cite{Kena-Cohen2010,Ramezani2017Plasmon}, enhanced exciton
	conductivity~\cite{Orgiu2015}, light harvesting~\cite{Coles2014Strong} and suppression of
	photo-bleaching in J-aggregates 	\cite{Munkhbat2018}. The strong light-matter coupling
	regime can be attained experimentally in a variety of different setups, such as in
	microcavities filled with a large number of molecules~\cite{Lidzey1998} or, more
	recently, in nanoscale plasmonic resonators coupled to just a few
	molecules~\cite{Zengin2015, Chikkaraddy2016}.
	
			The influence of strong light-matter coupling on molecular properties has recently come
	into focus, with several works predicting a reduction of vibrational effects in
	exciton-polariton states compared to bare-molecule excited states. This reduced
	vibrational displacement (RVD) could have significant consequences for the chemical
	reactivity of such molecules, permitting polaritonic-based catalysis of
	electron-transfer~\cite{Herrera2016} and photo-isomerisation~\cite{Galego2016}.
	Approaches typically rely on the Holstein-Tavis-Cummings (HTC) model, which describes a
	single EM mode coupled to a collection of molecules with a single vibrational mode
	each~\cite{Kirton2013,Cwik2014,Spano2015,Herrera2016,Zeb2016}.
	
	 Nevertheless, actual molecules 
	possess a large number of nuclear oscillation modes and additionally, interact with a
	continuous bath of phononic modes in the host solvent~\cite{George2015Liquid-Phase} or
	polymer matrix~\cite{Hakala2009} in typical experiments, which induces additional
	chemical changes~\cite{Nitzan2006}. In this case, the persistence of the RVD in
	polaritons is unknown.
	
			In this work, we present a study of the lowest-energy polaritonic state supported by a
	small ensemble of organic molecules under strong coupling to an optical micro-cavity
	mode, sketched in \autoref{Fig:fig1}a. Under the assumption that the molecules are close
	to their mechanical equilibrium, we include the whole vibrational spectrum of nuclear
	and environmental modes in our analysis, specified by the spectral density
	$J_v(\omega)$. The level of modelling of the molecules, therefore, goes beyond the HTC
	Hamiltonian, but still enables exploration of the full many-body wavefunction employing
	canonical bosonic operators and thus the toolbox from quantum optics. Due to its mixed
	threefold photonic-excitonic-phononic character, this eigenstate is dubbed the
	\textit{lower polaron-polariton} (LPP), adopting the nomenclature in
	ref.~\onlinecite{Wu2016}. To treat phononic and photonic processes in the molecules in
	the same footing, and, at the same time, deal with an arbitrary number of vibrations, we
	exploit a Tensor Network (TN) representation of the system's wavefunction that extends
	the widespread Matrix Product States (MPS) for one-dimensional quantum
	chains~\cite{Orus2014}. By this means, the LPP is retrieved by variational minimisation
	of the TN via the Variational Matrix Product State (VMPS)
	algorithm~\cite{Weichselbaum2009}, which filters out the most relevant components mixed
	in the state capturing their mutual entanglement.
	
		Our results show the robustness of RVD in the LPP wavefunction beyond the HTC model,
	extending its validity to arbitrary structured phononic baths. Moreover, the trend for
	larger ensembles indicates the effective suppression of polaron formation in the
	thermodynamic limit. Intriguingly, we observe that excitonic and photonic components are largely independent of
	specific molecular details, and can be reproduced by an effective HTC model determined by two
	single-molecule cumulative parameters.  Specifically, these are the
	\textit{reorganisation energy} $\Delta$~\cite{Marcus1956}, associated with the
	re-equilibration of the vibrational modes after electronic excitation, and the mean
	phononic oscillation frequency $\Omega_v$ that corresponds to a collective
	\textit{reaction coordinate}. Conversely, the LPP vibrational properties are strongly
	molecule-dependent and thus shaped by $J_v(\omega)$.
	
	The paper is organised as follows: in \autoref{sec:theory} we first present the model and
	introduce the TN-based algorithm aimed to target the LPP wavefunction. Namely, we analyse
	in \autoref{sec:sys_LPP} the LPP eigenfrequency and the excitonic and photonic states reduced
	populations as a function of vibronic coupling, including the vibration-free polaritonic
	components mixed into the state. In \autoref{sec:vib_LPP} focus on the `spectrum' of
	vibrational displacement as the spectral density is varied. Finally, we test previous
	conclusions for a particular realisation of organic polaritons (Rhodamine 800 molecule)
	in \autoref{sec:rhod800_GS}.
	
	\section{Model and methods}\label{sec:theory}
	
	\begin{figure}[tb]
		\includegraphics[width=\linewidth]{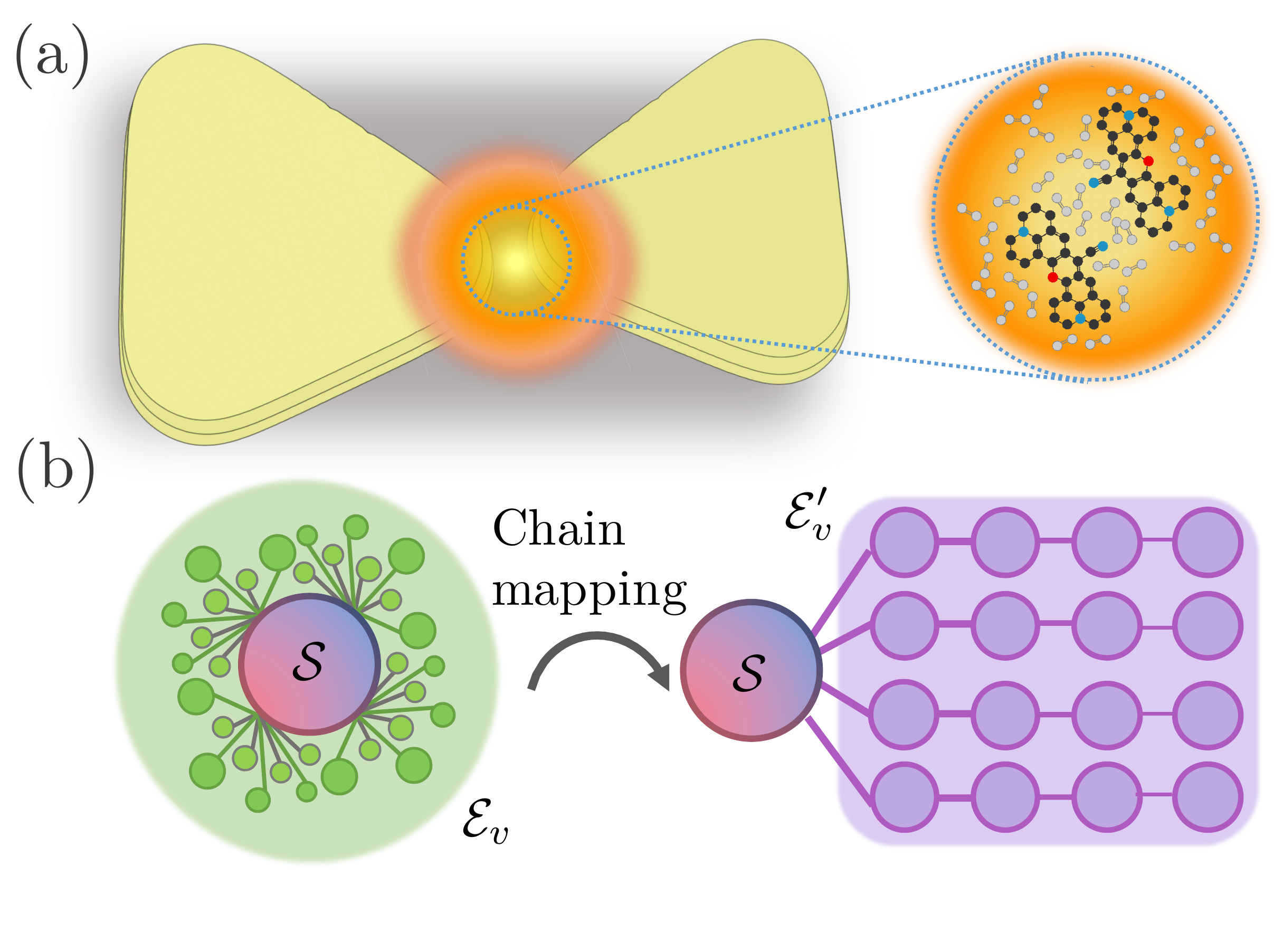}
		\caption{(a) Sketch of a molecular ensemble interacting with a confined EM resonance
	(dashed region) and with the host environment (grey circles). (b) Scheme illustrating
	the mapping of the vibrational modes yielding the multi-chain Hamiltonian used in the
	simulations, where the excitons interact with collective reaction coordinates in the
	molecules.}\label{Fig:fig1}
	\end{figure}
	Our model includes a collection of $N$ identical molecules, containing a single exciton
	with frequency $\omega_e$ and ladder operators $\hat{\sigma}_{\pm}^{(i)}$
	($i=\big(1,\cdots,N\big)$), placed within the volume of a resonant EM mode (frequency
	$\omega_O=\omega_e$) and annihilation operator $\hat{a}$.  The total Hamiltonian contains
	two different parts, as schematically depicted in \autoref{Fig:fig1}b. First, the
	system $\mathcal{S}$ that accounts for the excitons within the molecules, the cavity EM
	mode and their mutual coupling, measured by the collective Rabi frequency $\Omega_R$ and
	treated within the rotating-wave approximation (setting $\hbar=1$), 
	\begin{equation}
	\hat{H}_{\mathcal{S}}= \omega_O\hat{a}^{\dagger}\hat{a}+\sum_{i=1}^{N} 
	\omega_e \hat{\sigma}^{(i)}_{+}\hat{\sigma}^{(i)}_{-} + \frac{\Omega_R}{2\sqrt{N}}
	\sum_{i=1}^{N}(\hat{a}^{\dagger}\hat{\sigma}^{(i)}_-+\hat{\sigma}^{(i)}_+\hat{a}) \label{eq:Our_H}.
	\end{equation}
	We neglect inter-excitonic interactions, which we assume are screened out by the
	host environment. In the single-excitation subspace, $\hat{H}_{\mathcal{S}}$ is
	exactly solvable. Its eigenstates are two polaritons, upper (UP) and lower (LP),
	$\ket{\pm}= (\hat{a}^{\dagger}\ket{G} \pm \ket{B})/\sqrt{2}$, with frequencies
	$\omega_{\pm}=\omega_O\pm \Omega_R/2$, which result from the hybridisation of
	the collective excitonic \emph{bright} state $\ket{B}=(\sum_{i=1}^N
	\hat{\sigma}_+^{(i)}\ket{G})/\sqrt{N}$ with the cavity EM mode (here $\ket{G}$
	stands for the global vacuum state). In addition, there are $(N-1)$ so-called
	\emph{dark} states (DS), $\ket{d}\in \mathcal{D}$, which are purely excitonic states
	of frequency $\omega_e$ that are orthogonal to $\ket{B}$.
	
		The second part of the Hamiltonian describes the vibrational subspace
	$\mathcal{E}^{(i)}_v$ containing a large number $M_v$ of vibrational modes both inside
	the molecule and in the host environment, and their elastic coupling to the excitons. The
	$k$-th vibrational mode is approximated by a harmonic oscillator of frequency $\omega_k$
	(valid close to the equilibrium position) with annihilation operator $\hat{b}^{(i)}_{k}$
	and exciton-phonon coupling strength $\lambda^{(i)}_{k}$,
	\begin{equation}
	\hat{H}_{v} = \sum_{i=1}^{N} \sum_{k=1}^{M_v} \omega^{(i)}_{k} \hat{b}_{k}^{\dagger (i)}\hat{b}^{(i)}_{k} \, +
	\sum_{i=1}^{N} \sum_{k=1}^{M_v} \lambda^{(i)}_{k} (\hat{b}^{(i)}_{k}+\hat{b}_{k}^{\dagger (i)}) 
	\hat{\sigma}^{(i)}_{+}\hat{\sigma}^{(i)}_{-} .\label{eq:mol_H}
	\end{equation} 
	The properties of these modes, $\{\omega^{(i)}_k,\lambda_k^{(i)}\}$, are encoded in the
	spectral density $J_v^{(i)}(\omega) = J_v(\omega) = \pi\sum_{k=1}^{M_v}
	\lambda_k^2\delta(\omega-\omega_k)$.  Here the exciton-phonon coupling does not lead to
	nonradiative decay~\cite{Nitzan2006}. The coupling parameters $\lambda_{k}$ describe the
	relative shifts of electronic potential surfaces between ground and excited states. For
	the uncoupled molecules ($\Omega_R=0$), the total Hamiltonian
	$\hat{H}_\mathcal{S}+\hat{H}_v$ becomes a sum of exactly diagonalisable
	independent-boson-models~\cite{Mahan2000}. Namely, the eigen-excitations or Lang-Firsov
	polarons can be visualised as localised `phonon clouds' surrounding the excitons at each
	molecule: $\ket{D_i}=e^{-\hat{S}_i}\ket{G}$, with $\hat{S}_i = \hat{\sigma}_+^{(i)}
	\hat{\sigma}_-^{(i)} \sum_k \frac{\lambda^{(i)}_k}{\omega^{(i)}_k} (\hat{b}_{k}^{(i)} -	\hat{b}_{k}^{(i)\dag})$.
	
		For typical molecules, when $M_v$ becomes large (or even formally infinite when a
	continuum approximation for $J_v(\omega)$ is used), most standard approaches of quantum
	optics to calculate the eigenstates of the system fail, and even many approximate methods
	become prohibitively expensive. For instance, direct diagonalisation of the total
	Hamiltonian is unfeasible, even for $N=1$.  However, a VMPS algorithm permits to
	calculate the full wavefunction including all degrees of freedom of either the
	time-evolved wavefunction after excitation (analysed in~\onlinecite{DelPino2018}) or the
	ground state within the single electronic excitation subspace. This is precisely the
	`lower polaron-polariton' ~\cite{Spano2015,Herrera2016,Wu2016} analysed in the so-called
	Holstein-Tavis-Cummings (HTC) model~\cite{Spano2015,Herrera2016,Zeb2016,Wu2016}, which
	deals with a single vibrational mode with frequency $\omega_v$ and vibronic coupling
	$\lambda_v$, i.e., $J_{\mathrm{HTC}}(\omega)=\pi\lambda_v^2 \delta(\omega-\omega_v)$. In
	particular, the Hamiltonian \autoref{eq:mol_H} extends the HTC to an arbitrary number
	$M_v$ of vibrational modes.
	
		Here we employ a Tensor Network State-based approach that is numerically exact as long as
	convergence is reached, permitting a nonperturbative exploration of the quantum effects
	arising in the LPP\@. It does not rely on any specific form of the spectral density, and
	in particular, can be used both for discrete and continuous $J_v(\omega)$.     To apply
	this method, we perform an orthogonal mapping of the modes in the $N$ vibrational (green in \autoref{Fig:fig1})
	environments ($\mathcal{E}^{(i)}_v$). We rely on the chain transformation introduced
	in~\cite{Prior2010}, which maps the molecular Hamiltonian \autoref{eq:mol_H} for an
	exciton coupled to many independent vibrations to an exciton coupled only to the first
	site in a chain of coupled oscillators, leading to the vibrational Hamiltonian
	$\hat{H}'_v=\sum_{i=1}^N\hat{H}_v^{'(i)}$
	\begin{multline}\label{eq:H_chain}
	\hat{H}_v^{'(i)} = \Omega_v\hat{c}_{0}^{(i)\dagger}\hat{c}^{(i)}_{0}+\eta\hat{\sigma}_+^{(i)}\hat{\sigma}_-^{(i)}\big(\hat{c}^{(i)}_{0}+\hat{c}_{0}^{(i)\dagger}\big) +\\
	\sum_{l=1}^{M_v-1}\omega_l\hat{c}_{l}^{(i)\dagger}\hat{c}^{(i)}_{l}+ \sum_{l=0}^{M_v-2}t_l\big(\hat{c}^{(i)\dagger}_{l}\hat{c}^{(i)}_{l+1}+\hat{c}^{(i)\dagger}_{l+1}\hat{c}^{(i)}_{l}\big).
	\end{multline}
	This transformation results in the \textit{star} coupling structure for $\hat{H}_\mathcal{S}+\hat{H}'_v$ sketched in the right part of \autoref{Fig:fig1}b. The modes are thus regrouped in chains with length
	$L=M_v$~\cite{Chin2011}, with only the first mode coupled to the exciton-photon
	subspace $\mathcal{S}$ (red-blue). Namely, each exciton is coupled to a single collective
	reaction mode defined by $\eta \hat{c}_0 = \sum_k \lambda_k    \hat{b}_k$~\cite{Garg1985,Thoss2001,Renger2002,Iles-Smith2014}, with total amplitude 
	\begin{equation}\label{eq:eta}
	\eta=\sqrt{\int_0^{\omega_c}J_v(\omega)\hspace{1mm}\mathrm{d}\omega/\pi},
	\end{equation}
	which generalises the vibronic coupling in the HTC model ($\eta=\lambda_v$ in this case), and average bath frequency 
	\begin{equation}
	\Omega_v=\frac{\int_0^{\omega_c}\omega J_v(\omega)\hspace{1mm} \mathrm{d}\omega}{\int_0^{\omega_c} J_v(\omega)\hspace{1mm} \mathrm{d}\omega},
	\end{equation}
	which similarly generalises $\omega_v$.   All other chain modes become connected through 
	nearest-neighbor hopping interactions $t_l$. The discussion below shows an important quantity to characterise the vibronic coupling in the system is the so-called reorganisation energy, given by
	\begin{equation}\label{eq:reorg}
	\Delta = \frac{1}{\pi}\int_0^{\omega_c} \frac{J_v(\omega)}{\omega} \mathrm{d}\omega,
	\end{equation}
	and directly linked with the Stokes (frequency) shift between maxima of emission and absorption spectra in organic
	molecules, given by $2\Delta$~\cite{Coles2010,Shreve2007}.
	
		The star Hamiltonian \autoref{eq:H_chain} permits direct implementation of the VMPS
	algorithm. To this end, the LPP wavefunction $\ket{\psi_-}$ is represented by a tensor
	network with maximum bond dimensions $D$, which directly mimics the coupling structure in
	the star Hamiltonian (\autoref{Fig:fig1}b), making the representation numerically
	efficient. This procedure, coupled with a variational approach to calculate the LPP,
	leads to a multi-chain variant of the Density Matrix Renormalisation Group algorithm
	(DMRG) for 1D quantum lattices~\cite{White1992}. When
	continuous vibrational spectra are taken into account, the formally infinite phononic
	chains are truncated at length $L$, which must be chosen large enough to reach
	convergence. A more detail discussion can be found in the Appendix, \ref{sec:TN_approach} and \ref{sec:DMRG_alg}.
	
			We present results for a maximum of $N=5$ molecules. For large molecular ensembles, a
	severe memory bottleneck occurs if the system $\mathcal{S}$ is stored in a single root
	tensor leading to exponential scaling in $N$. However, as shown
	in~\cite{Szalay2015,Shi2006}, it would be possible to restore the efficiency of the
	numerical method while maintaining precision, by further decomposing the root node into
	a tree TN, where each final branch represents an exciton or the cavity photon and is
	coupled to a single chain.  Developing a suitably efficient tree model requires an
	explicit analysis of `entanglement topology' of the state, an idea that has recently
	been implemented to allow the simulation of multi-environment linear vibronic models
	constructed from ab initio parametrisations of small molecules~\cite{Schroder2017}. This
	is precisely the approach followed to simulate time dynamics of this system
	in~\cite{DelPino2018}. To further ameliorate memory issues for large chain mode
	occupations, we employ an optimal boson basis for the chain tensors~\cite{Guo2012},
	which can be determined on the fly via VMPS~\cite{Schroder2016}. A more detailed
	description of the theoretical approach can be found in~\cite{Schroder2016}.
	
	\subsection{The Holstein-Tavis-Cummings limit}\label{sec:HTC}
	\begin{figure}[tb]
		\centering
		\includegraphics[width=\linewidth]{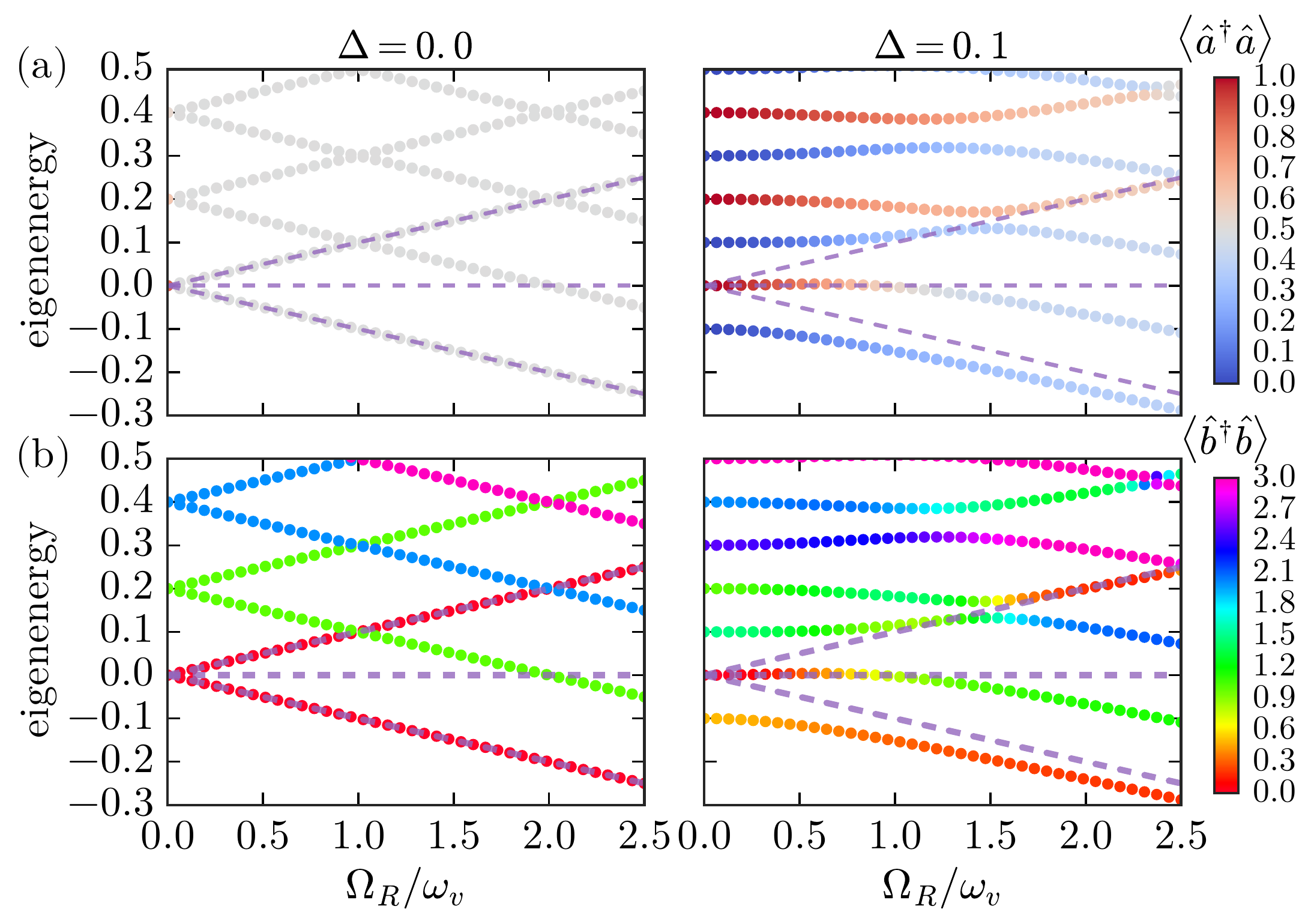}
		\caption{(a) Photonic components in the eigenspectrum of the HTC model for a single molecule as a function of the ratio between Rabi frequency $\Omega_R$ and vibrational frequency $\omega_v$ (b) Vibrational component. Dashed purple lines indicate the upper and lower vibration-free polariton energies. The energy origin is set at $\omega_e=\omega_{O}=0$.}\label{Fig:HTC_single}
	\end{figure}
	
	\begin{figure}[tb]
		\centering
		\includegraphics[width=\linewidth]{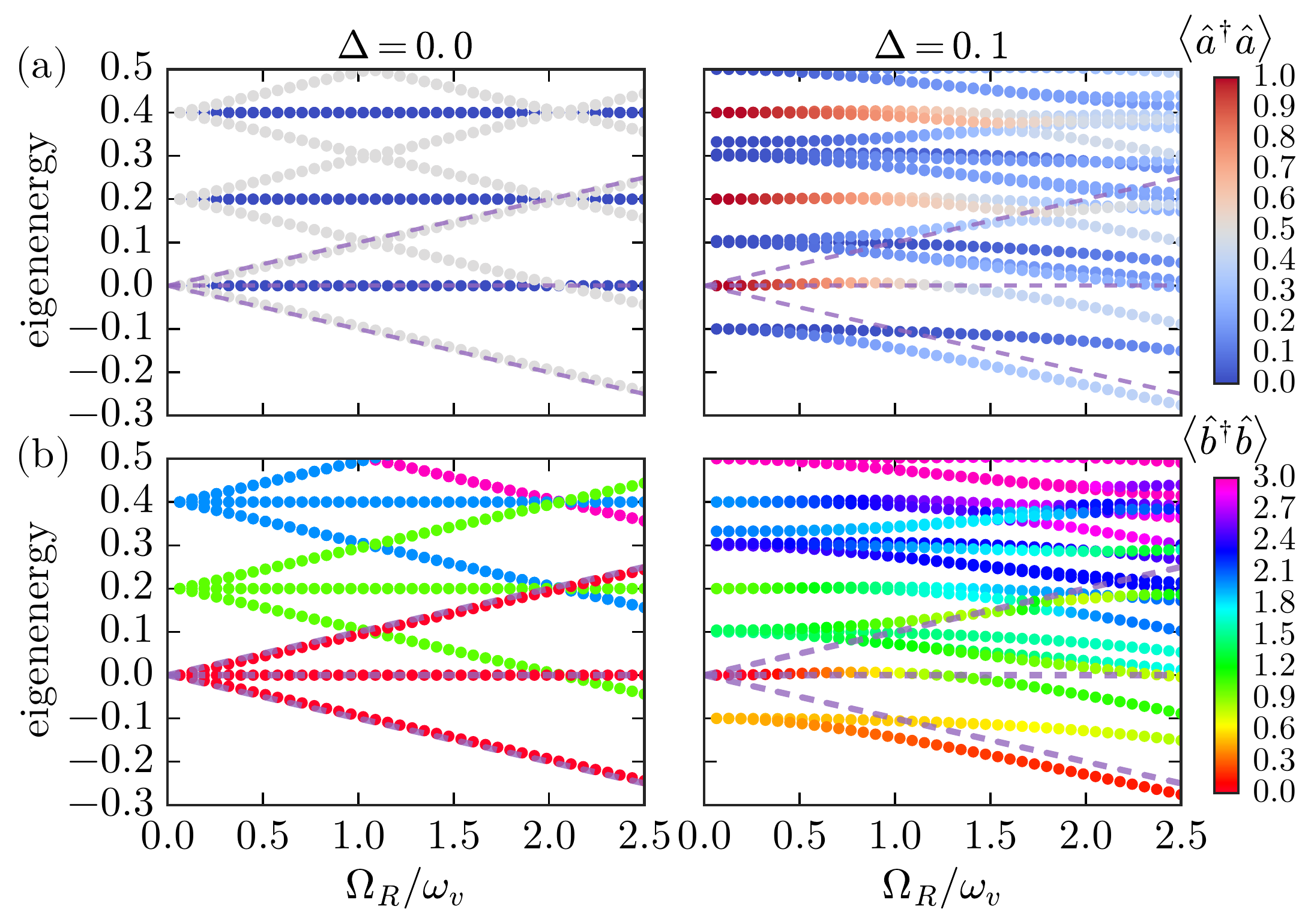}
		\caption{(a) Photonic components in the eigenspectrum of the HTC model for two molecules as a function of the ratio between Rabi frequency $\Omega_R$ and vibrational frequency $\omega_v$. (b) Total vibrational components, $\braket{\hat{b}^\dagger\hat{b}}=N\braket{\hat{b}^{(i)\dagger}\hat{b}^{(i)}}$. Dashed purple lines indicate the upper and lower vibration-free polariton and excitonic dark state energies. The energy origin is set at $\omega_e=\omega_{O}=0$, and we consider $\omega_v=0.2$ eV.}\label{Fig:HTC_two}
	\end{figure}
	
			In order to connect our study with the cases available in the literature, we start by
	briefly revisiting the HTC model, first introduced by Kirton et al.~\cite{Kirton2013},
	which can be solved by various methods, including direct numerical diagonalisation and
	via variational ansätze~\cite{Herrera2016, Zeb2016, Wu2016}. Vibronic coupling is
	parametrised by the reorganisation energy $\Delta=\lambda_v^2/\omega_v$, obtained from
	\autoref{eq:reorg}. In \autoref{Fig:HTC_single} and \autoref{Fig:HTC_two} we show the
	eigenenergies of the system for $N=1$ and $N=2$ molecules, with a single
	electronic/optical excitation in the system at maximum. The coexistence of vibronic and
	photonic couplings results in an involved eigenspectrum where states are characterised
	by a triple mixture of vibrational, photonic and excitonic
	states~\cite{Herrera2016,Herrera2017}.  In these figures, the color-scale is used for
	displaying the cavity and vibrational populations, $\braket{\hat{a}^{\dagger}\hat{a}}$
	and $\braket{\hat{b}^{\dagger}\hat{b}}=\sum_i\braket{\hat{b}^{(i)\dagger}\hat{b}^{(i)}}$, respectively.
		
			For a single molecule, the `vibration-free' eigenstates (i.e., for $\Delta=0$) system
	are precisely the polaritons described in left panels of \autoref{Fig:HTC_single},
	together with their vibrational sidebands corresponding to excited molecular phonons.
	The energies of these states are, therefore, $\omega_n^{\pm}=\omega_{\pm}+n\omega_v$,
	$n\in\mathbb{N}$. Conversely, if there is no coupling with the cavity mode (limit
	$\Omega_R=0$),  vibronic coupling results in the formation of a polaron with energy
	$\omega_e-\Delta$. For intermediate energy scales (right panels), the ground state in
	the single excitation space (LPP) has a partially polaritonic nature~\cite{Wu2016}. In
	\autoref{Fig:HTC_single}a, higher-energy polaritonic sidebands display anti-crossings
	caused by vibronic interaction when resonance conditions are met. This is confirmed by
	the fact that eigenstates do not have a well-defined phonon number in the vicinities of
	the split regions (see \autoref{Fig:HTC_single}b). Also, for large $\Delta$, the
	photonic nature of the states becomes transferred to higher energy excitations as it
	becomes increasingly unfavourable to form hybrid light-matter quasiparticles in the
	system, while for higher $\Omega_R$ the states contain a greater fraction of the
	`original' polaritons. As a minimum extension to the collective behaviour of the system,
	we analyse in the following a molecular dimer within the HTC model. In this case, the
	presence of one dark excitonic state
	$\ket{d}=(\hat{\sigma}_+^{(1)}-\hat{\sigma}_+^{(2)})\ket{G}/\sqrt{2}$ besides
	vibrational sidebands (frequencies $\omega_d^n=\omega_e+n\omega_v$, $n\in\mathbb{N}$) 
	is also present in the vibration-free eigenstates displayed in the left panels of
	\autoref{Fig:HTC_two}. Vibronic interaction results in a larger shift of dark states as
	compared to polaritons, as inferred from the right panels in \autoref{Fig:HTC_two}.
	Eigenenergies present actual crossings in addition to anticrossings, leading to a much
	richer structure of eigenstates than for $N=1$~\cite{Herrera2017}. Comparing the
	vibrational components in \autoref{Fig:HTC_two}b with those of \autoref{Fig:HTC_single}b
	above, a slight reduction of the vibrational dressing (RVD) in the LPP is noticed, 
	close to the threshold at which reorganisation energy and Rabi frequency become
	comparable (i.e., when $\Omega_R\simeq\Delta$).
	
		In particular, numerical and variational solutions for larger $N$ display scaling of this
	phenomenon as $1/N$ \cite{Herrera2016,Galego2016,Zeb2016}. In the thermodynamic limit,
	therefore, the LPP becomes closer to the bare LP, and polaron formation is suppressed. A
	key observation is that RVD is not observed for the whole ladder of eigenstates,
	resulting in strong vibronic effects arising in the excited states (for example in the
	dynamics triggered by a high-frequency pulsed excitation). One direct consequence of
	polaron decoupling is the eventual suppression of the reorganisation energy of excited
	electrons, which, as first pointed by \cite{Herrera2016} could be exploited to enhance
	electron transfer reaction rates.

	\section{Excitonic and photonic features of the LPP}\label{sec:sys_LPP}
	
	The spectral density in organic materials depends strongly on the molecules and the host
	matrix. In order to obtain general conclusions, we thus first study the effect of
	many-mode vibrational dressing on polaritons using a standard parametrisation of
	low-frequency vibrational modes, the Leggett-type spectral density\footnote{Spectral densities of this
	type have been studied thoroughly in the simplest example of quantum dissipation model,
	the Spin-Boson Model. See~\cite{Leggett1987} and~\cite{Weiss1999} for further details.}
	\begin{equation}\label{eq:Leggett}
	J_v(\omega) = 2\pi\alpha\omega_c^{1-s}\omega^s\theta(\omega_c-\omega), \qquad s>0\,,
	\end{equation}
	where $\omega_c$ corresponds to a cutoff energy, $\alpha$ describes the overall strength
	of the exciton-phonon coupling and the exponent $s$ determines the shape of the spectral
	density, with $s=1$ giving a so-called `Ohmic' spectral density, while $s<1$ and $s>1$
	correspond to sub- or super-Ohmic densities, respectively. The cutoff $\omega_c$ implies
	that fast modes $\omega_k>\omega_c$ are reabsorbed in the coupling constants through the
	adiabatic approximation~\cite{Weiss1999}. It is interesting to note for the following
	discussion that, within the single molecule limit, a regime  recently reported in
	plasmonic nanocavities~\cite{Zengin2013,Chikkaraddy2016}, the Hamiltonian
	\autoref{eq:mol_H} can be mapped precisely to the well-known Spin-Boson Model
	(SBM)~\cite{Leggett1987, Weiss1999} by a shift of the vibrational mode origin, as shown
	in \autoref{sec:single_mol} of the Appendix.
	
	\begin{figure}[tb]
		\centering
		\includegraphics[width=\linewidth]{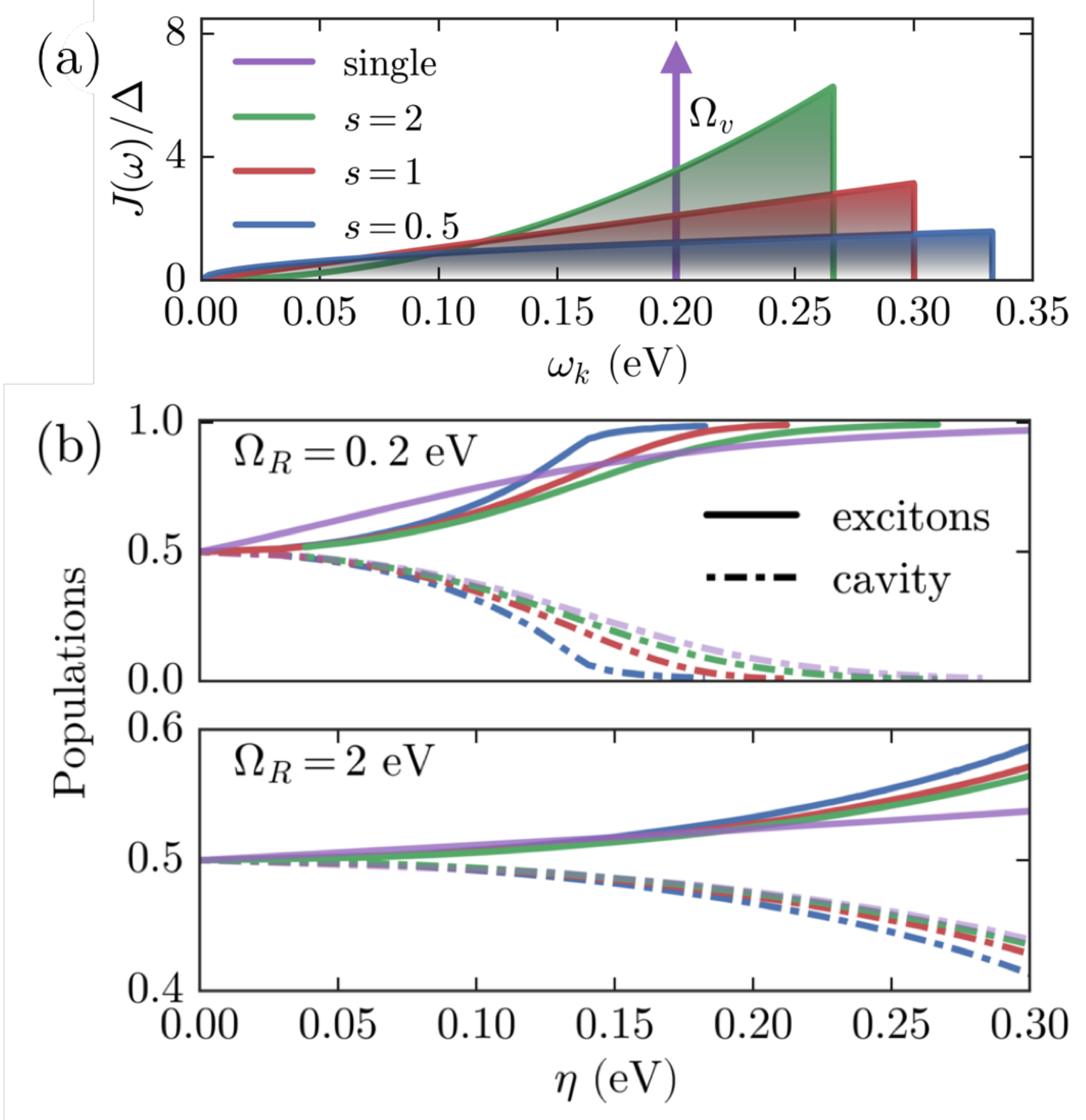}
		\caption{(a) Functional forms adopted for the LPP
			calculation and (b) Reduced populations over subsystem $\mathcal{S}$ in the LPP as a function of the coupling strength to the reaction coordinate $\eta$. Here different curves correspond to the different spectral densities in (a), depicting in the upper and lower windows of (b) the cases for Rabi frequency shown.}\label{Fig:fig2}
	\end{figure}
	
	\begin{figure*}[tb]
		\includegraphics[width=\linewidth]{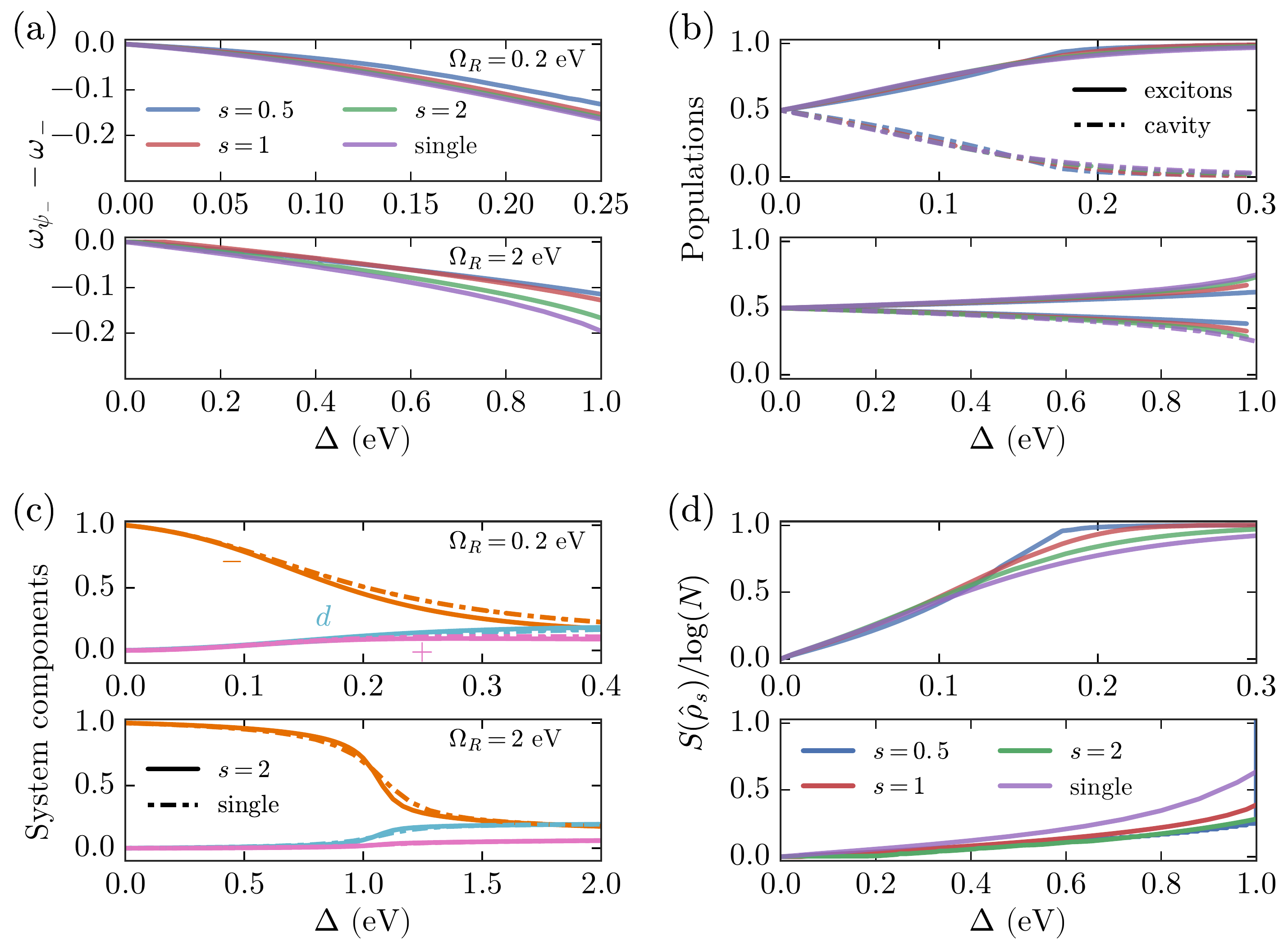}
		\caption{(a) Energy shift with respect to the bare LP
			($\Delta=0$), (b) exciton/photon fractions (c), vibration-free components in the LPP and (d) Von-Neumann entropy of the
			photonic-excitonic and vibrational states bipartition, as a function of reorganisation energy. In these panels upper panels depict the case $\Omega_R=\Omega_v$ and
			$\Omega_R=10\Omega_v=2$ eV for the lower panels.}\label{Fig:fig3}
	\end{figure*}
	
			In this section, we study the influence of vibrational dressing on the excitonic and
	photonic properties of the LPP. We focus on Leggett-type spectral densities,
	\autoref{eq:Leggett}, with $s=0.5$, $s=1$, and $s=2$ (shown in \autoref{Fig:fig2}a). For
	ease of reference, we will compare the results with those obtained from the HTC model.
	The cut-off frequency $\omega_c$ is tuned to maintain the reaction coordinate frequency
	constant and equal to that of the reference HTC model $\Omega_v = \omega_c (1+s)/(2+s)$
	($=\omega_v$ in the HTC). We consider an anthracene-like molecule having a vibrational
	spacing of $\Omega_v=0.2$ eV, and unless otherwise stated, we assume an ensemble
	containing $N=5$ molecules.
	
	In order to investigate the interplay between exciton-photon and exciton-phonon
	interactions without dealing with the full many-body state, we trace out the vibrational
	modes to calculate the reduced density matrix,
	$\hat{\rho}_\mathcal{S}=\mathrm{Tr}_{\mathcal{E}_v}\{\ket{\psi_-}\bra{\psi_-}\}$.    As
	\autoref{Fig:fig2}b shows, the LPP state becomes more excitonic as the coupling amplitude
	to the reaction coordinate $\eta$ \autoref{eq:eta} grows, with a more substantial
	photonic character in a given molecule if the Rabi frequency $\Omega_R$ is larger. Such
	behaviour signals  the trade-off between vibrational and photonic coupling scales.
	Nevertheless, the different curves for different $J_v(\omega)$ clearly indicate that the
	vibrational influence on the system is not solely determined by the reaction coordinate
	in the organic molecules. Instead, the remaining `dark' vibrational combinations produce
	changes that are not captured adequately by $\eta$.
	
		These results motivate us to consider the reorganisation energy $\Delta$, defined by
	\autoref{eq:reorg} and given by $\Delta = 2\alpha\omega_c/s$ for Leggett-type
	$J_v(\omega)$, as a measure of the global influence of the whole set of phononic modes on
	the system. We next proceed to include the LPP eigenfrequency $\omega_{\psi_-}$ in the
	discussion. In particular, the values of the shift $\omega_{\psi_-}-\omega_-$ (the bare
	LP frequency to disregard the linear energy shift by the Rabi frequency in the
	discussion) as a function of the reorganisation energy can be read from
	\autoref{Fig:fig3}a. The LPP eigenenergy undergoes a monotonic red-shift as a result of
	the increased vibrational dressing in electronic and photonic components. Mirroring the
	results for the HTC model, the slope of the curves differs at either side of the
	crossover at $\Delta\simeq\Omega_R/2$, being steeper at the large $\Delta$
	region~\cite{Zeb2016}. Accompanying this trend, the total photonic (excitonic) fractions
	of the LPP, (diagonal elements in $\hat{\rho}_\mathcal{S}$) are decreasing (increasing)
	very similar functions of the reorganisation energy (\autoref{Fig:fig3}b). Additional
	insight into the LPP internal structure is gained by considering the weights of the
	vibration-free eigenstates ($\Delta=0$) that are mixed into the LPP by the vibronic
	coupling. As expected, \autoref{Fig:fig3}c reveals that $\ket{\psi_-}$ is constituted by
	major contributions from the bare LP ($\ket{-}$) and dark states
	($\ket{d}\in\mathcal{D}$), with some small fraction of the UP ($\ket{+}$). In particular,
	in the flatter region where $\Delta\ll\Omega_R/2$ the LPP is well approximated by the
	bare lower polariton with frequency $\omega_{\psi_-} \simeq \omega_- =
	\omega_e-\Omega_R/2$.
	
	The results above suggest that in real space the limit at large Rabi frequencies
	corresponds to a LPP that is a spatially delocalised state over the ensemble, with
	nearly no molecular phonon excitation. In the opposite large Stokes-shift limit, the LPP
	becomes closer to spatially localised polarons, with a LPP-LP shift of $-\Delta$
	(irrespective of $J_v(\omega)$), and no photonic component. The trade-off between the
	two effects leads to a relocation of the contribution $\rho_{--}$, into the dark
	excitonic states, $\rho_{\mathcal{D},\mathcal{D}}$, with eventual crossover at the large
	Stokes shift limit, mirroring results for the HTC model in the previous section. To
	further corroborate the existence of the previous extremes we calculate the bipartite
	entanglement between $\mathcal{S}$ and $\mathcal{E}$, measured by the partial von
	Neumann entropy $S(\hat{\rho}_\mathcal{S}) =
	-\mathrm{Tr}(\hat{\rho}_\mathcal{S}\log\hat{\rho}_\mathcal{S})$. As we observe in
	\autoref{Fig:fig3}d, correlations in the LPP, absent at $\Delta=0$, build up for
	increasing Stokes shift as vibrational states become more mixed with excitons (and
	indirectly the cavity photon). For large $\Delta$, the formation of polarons entails the
	saturation of the entanglement entropy at the value $S(\hat{\rho}_\mathcal{S})\simeq\log
	N$ corresponding to $N$ excitons maximally mixed with the vibrational modes.
	
	\begin{figure}[tb]
		\centering
		\includegraphics[width=\linewidth]{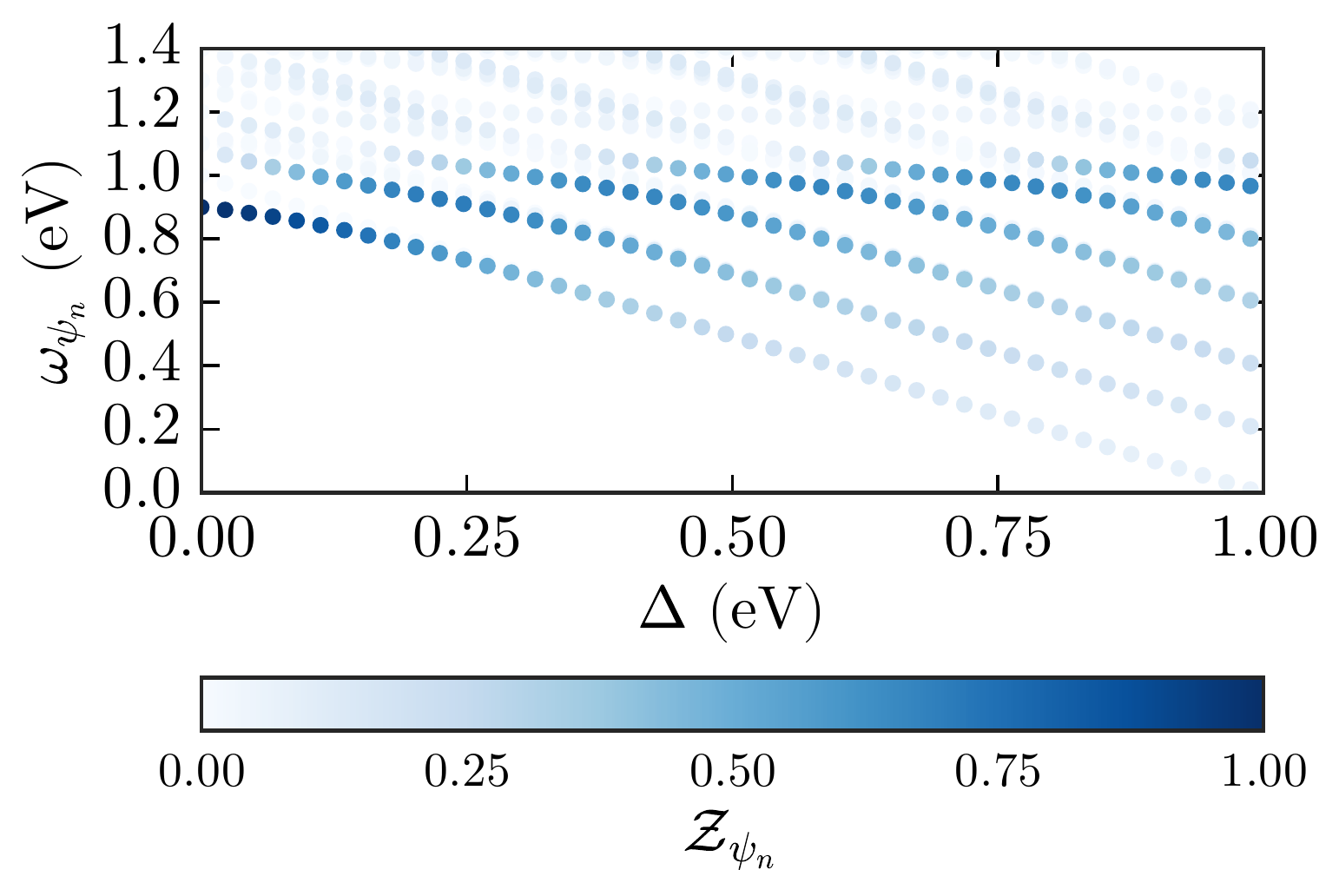}
		\caption{Excitation spectrum $\ket{\psi_n}$ of $\hat{H}_\mathcal{S}+\hat{H}_v$ for a HTC model with $N=2$ molecules and
			$\Omega_R=0.2$ eV. The color-scale indicates the overlap with the
			bare LP $\mathcal{Z}_{\psi_n}$ with increasing
			saturation.}\label{Fig:fig4}
	\end{figure}
	
			In the halfway of the polaritonic and polaronic limits, where neither $\Omega_R$ or
	$J_v(\omega)$ are negligible, state $\ket{\psi_-}$ possesses a threefold
	excitonic-photonic-polaronic character. Even in this region, observables acting within
	the system $\mathcal{S}$ present only quantitative changes depending on the vibrational
	spectrum of the molecules. A direct consequence is the emergence of an RVD effect when
	comparing the LPP with uncoupled excitons, arising from an increased Rabi frequency
	(lower panels in \autoref{Fig:fig3}). In this way, shallower curves in
	\autoref{Fig:fig3}a,b,c indicate a larger resemblance with the LP in frequency,
	populations and light-matter coherences respectively, while the entanglement entropy
	$S(\hat{\rho}_\mathcal{S})$ diminishes in \autoref{Fig:fig3}d as $\Omega_R$ is augmented
	due to decoupling from vibrations. In our simulations, we also observed similar effects
	for a fixed Rabi frequency and growing ensemble size $N$. The increasing suppression of
	the vibrational dressing could suggest the onset of polaron decoupling in the limit
	$N\gg1$,  similar to the one reported for a single mode as observed
	above~\cite{Herrera2016, Zeb2016}. From the point of view of chemistry, similar results
	have been obtained without canonical quantisation of molecular vibrations, generalising
	our conclusions to arbitrary electronic potential energy surfaces far from-equilibrium
	(e.g. in chemical reactions)~\cite{Galego2015,Galego2016}. In this case, the so-called
	`collective protection' (the extended version of the RVD) has been shown in the
	thermodynamic limit, by means of suppressed displacements in the polaritonic
	(nuclear-coordinate-resolved) surfaces. Lower polariton and ground state (vacuum)
	electronic surfaces thus become identical, with a critical impact on photo-isomerisation
	reactions.
	
	The results above reveal the robustness of the RVD, which does not rely on the fine
	molecular/host details encoded in $J_v(\omega)$.  Instead, the LPP for molecules close to
	the equilibrium is accurately characterised by the coupling to a reaction coordinate with
	frequency $\Omega_v$ while the global effect of the vibrational modes enters through the
	full reorganisation energy $\Delta$. These features can therefore be understood within an
	effective HTC model, with phonon frequency chosen according to the experimental reaction
	coordinate frequency and Stokes shift.
	
	The case of polaritons formed from molecules with large Stokes shift has been recently
	reported, showing that emission from the hybrid polaritons is suppressed and instead
	occurs from the bare molecules \cite{Baieva2017}. In this limit, it is more energetically
	favourable for an exciton to form a polaron rather than hybridise with a photon, as shown
	by diagonalisation of $\hat{H}_\mathcal{S}+\hat{H}_v$. Although TN-based approaches to
	target excited states have been devised~\cite{Dargel2012}, we exploit aforementioned
	similarities to further visualise the photonic properties of a simpler HTC model with
	$N=2$  (see \autoref{Fig:fig4}). In this manner we observe that the quasi-particle weight
	of the vibration-free polariton $\mathcal{Z}_{\psi_n}=||\braket{\psi_-|\psi_n}||^2$
	indeed `climbs up' to the excited states as $\Delta$ is increased, a fingerprint of
	polaritons at higher energies (see \autoref{Fig:fig4}). This can be understood as a
	consequence of the fact that transitions within the Franck-Condon region (and thus with
	non-negligible dipole moment) occur to vibrationally excited states for large
	exciton-phonon coupling.
	
\begin{figure}[tb]
		\centering
		\includegraphics[width=\linewidth]{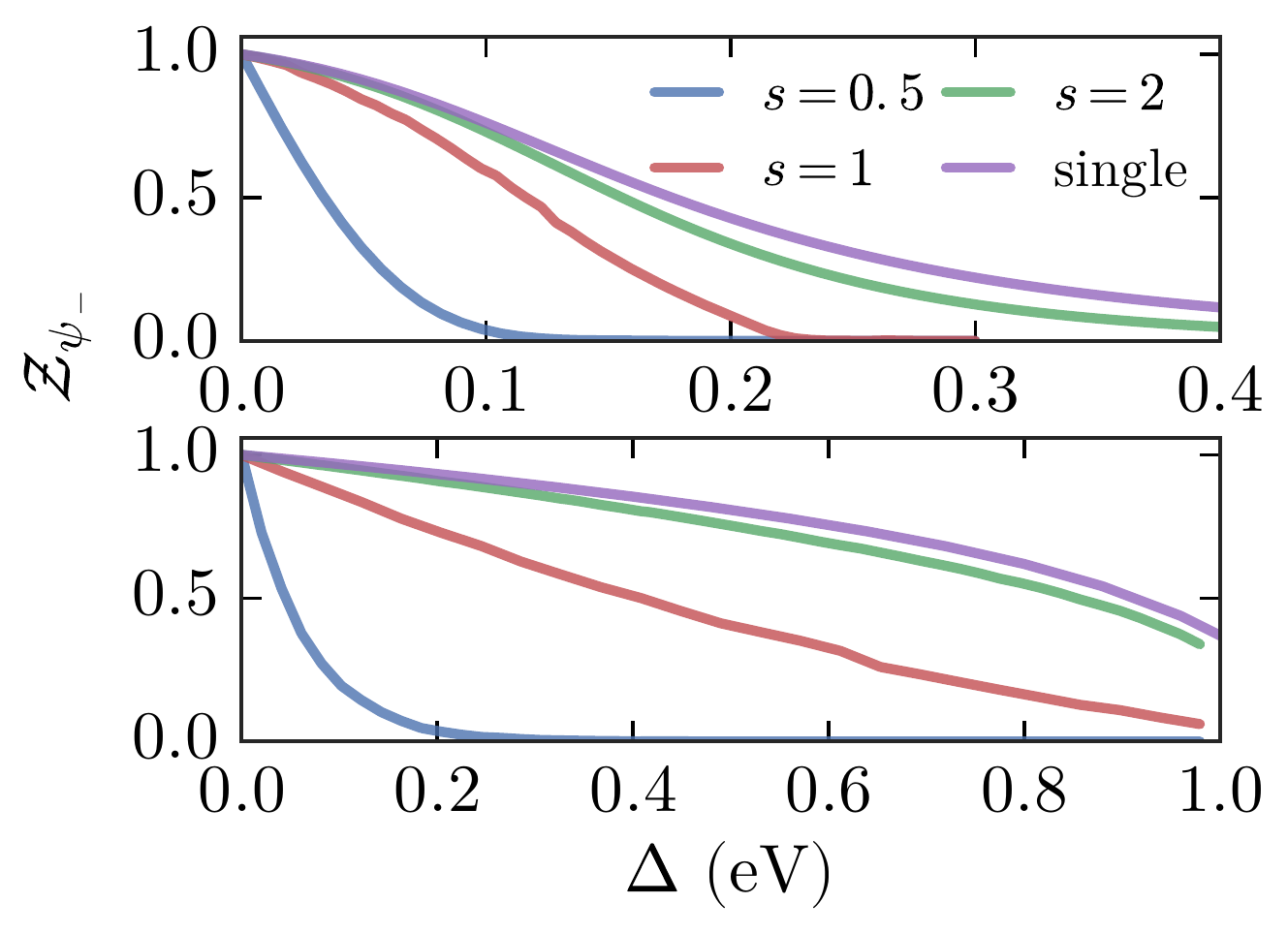}
		\caption{Overlap of the LPP wavefunction
			with the vibration-free polariton $\ket{-}$ for $\Omega_R=0.2$ eV (upper panel) and $\Omega_R=2$ eV (lower panel) for the HTC and $J_v(\omega)$ in the form \autoref{eq:Leggett}.}\label{Fig:fig5}
	\end{figure}

	Finally, we noticed quantitative differences as molecular details were varied, which are
	particularly prominent in the region $\Delta\simeq\Omega_R/2$, where maximum mixing
	between purely polaritonic and polaronic states is observed, and when the vibrational
	spectrum in the molecules is dominated by slow modes (sub-Ohmic $J_v(\omega)$). While
	exciton/photon properties have been shown to be universal, the impact of light-matter
	coupling on different molecular vibrations is averaged out in $\hat{\rho}_\mathcal{S}$.
	These non-trivial features motivate further analysis of the intrinsic vibrational
	properties in the LPP state, by first calculating a quasiparticle weight
	$\mathcal{Z}_{\psi_-}=\Vert\braket{\psi_-|-}\Vert^2$ that indicates how close the LPP is
	to the vibration-free LP, shown in \autoref{Fig:fig5}. Increased overlaps signal the RVD
	at large Rabi frequencies and exponents $s$. It should be noted that
	$\mathcal{Z}_{\psi_-}$ includes only the overlap with the vibrational vacua
	$\ket{0}_{\mathcal{E}^{(i)}_v}$, in contrast with the population $\rho_{--}$, reduced
	over the whole environment. Therefore, when \autoref{Fig:fig3}c and \autoref{Fig:fig5}
	are compared, a more polaronic character of the LPP is noticed with larger vibrational
	dressing for sub-Ohmic $J_v(\omega)$. However, as we show in  \autoref{Fig:fig3}a, the
	energy shifts associated with sub-Ohmic baths are actually weaker than in super-Ohmic and
	single mode environments, indicating that the RVD is strongest for sub-Ohmic baths. We
	shall elucidate the origin of this effect, related to the dominance of slow vibrational
	modes in sub-Ohmic environments, in the next section by further exploring the many-body
	vibrational properties of the LPP state.
		
	\section{Vibrational features of the LPP}\label{sec:vib_LPP}
	
		Our VMPS approach enables access to the full many-body vibrational component of the LPP
	wavefunction, which can be exploited to resolve phononic features in frequency space that
	are disregarded in the HTC model. In this section we show how vibrational observables are
	shaped by the specific shape of $J_v(\omega)$. To this end, we analyse the
	frequency-resolved vibrational displacement in the LPP state for the electronically
	excited molecule $i$. For further insight we split this observable into
	\textit{conditional displacement} contributions arising from each system state
	$\ket{n_\mathcal{S}}\in\mathcal{S}$ mixed into the LPP,
	\begin{equation}
	\mathcal{X}_{i,n_\mathcal{S}}(\omega_k,\Omega_R,\Delta)=\frac{|\braket{\psi_-|\hat{\mathcal{P}}_{n_\mathcal{S}}(\hat{b}^{(i)}_k+\hat{b}_k^{(i)\dagger})|\psi_-}|}{\rho_{n_\mathcal{S},n_\mathcal{S}}},\label{eq:disp_spectrum}
	\end{equation}
		which includes a projector over system states $\hat{\mathcal{P}}_{n_\mathcal{S}}$. Here
	we normalised by the corresponding system state population, to discern situations where
	the state fraction is vanishingly small.  In practice, the calculation of
	\autoref{eq:disp_spectrum} requires reverting the chain mapping for the $\hat{c}^{(i)}_l$
	modes.
	\begin{figure*}[tb]
		\includegraphics[width=\linewidth]{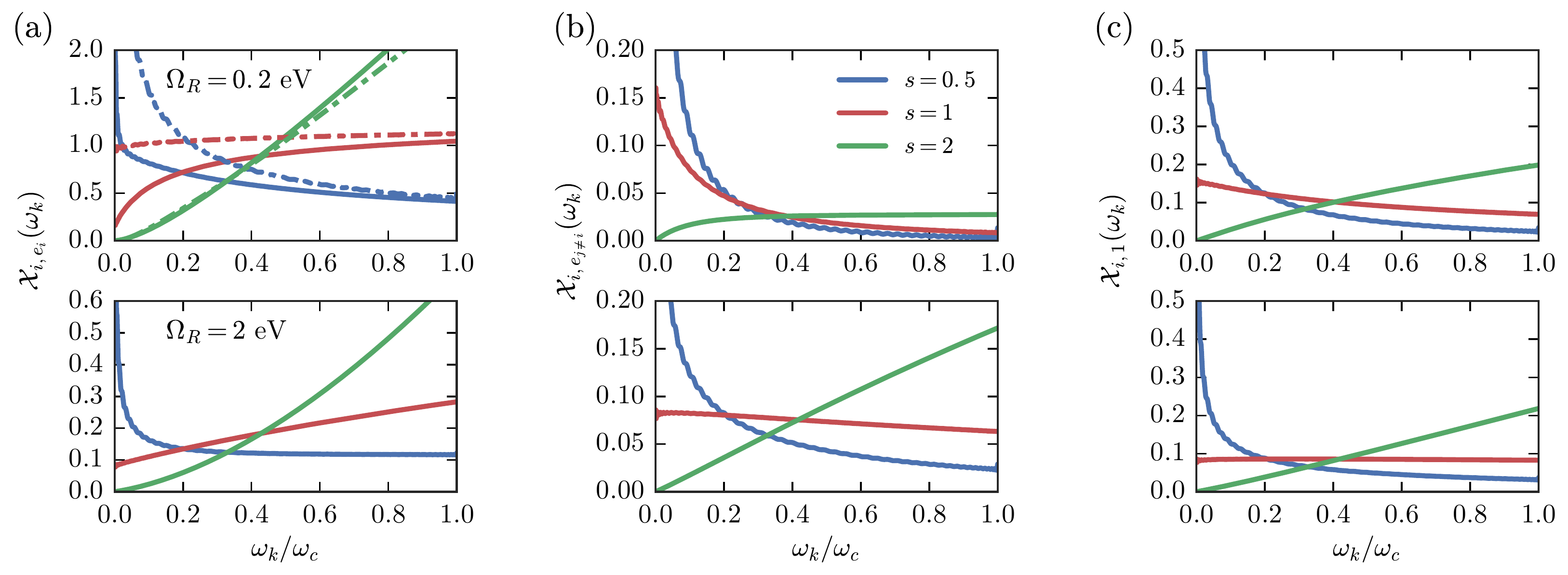}
		\caption{Absolute displacement spectrum of the phononic modes for a molecule $i$,
			projected on (a) the electronic state of the excited molecule,
			the electronic state of the unexcited molecule (b) and the
			cavity photon (c). In these panels we considered $\Omega_v=0.2$ eV, $\Delta=0.1$ eV and $N=5$. The limit $\Omega_R\rightarrow0$ is displayed in dashed-dotted lines.}\label{Fig:fig6}
	\end{figure*}
		The displacement spectra in \autoref{Fig:fig6} explicitly show the RVD of electronically
	excited molecules in the LPP , with decreased vibrational displacement
	$\mathcal{X}_{i,e_i}$ from the bare-molecule value
	$\mathcal{X}_{i,e_i}(\omega_k,\Omega_R=0) = \lambda_k^2/\omega_k^2$, when $\Omega_R$ is
	increased (see \autoref{Fig:fig6}a) and a more substantial suppression at low frequencies
	and smaller $s$.
	\begin{figure*}[tb]
		\centering
		\includegraphics[width=\linewidth]{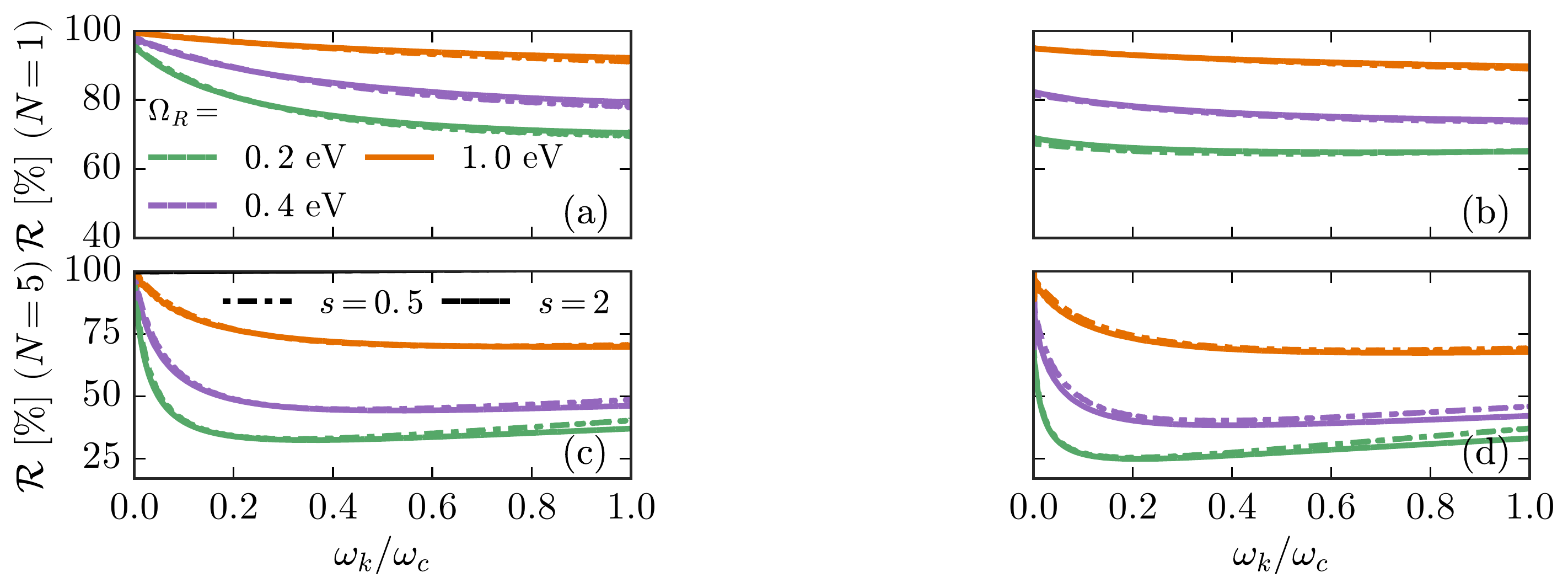}
		\caption{Ratio $\mathcal{R}$ between displacement at $\Omega_R'=2$ eV and
			lower Rabi frequency displacements, for the SBM at $N=1$ (a),(b),
			and $N=5$ (c),(d). The case $\Delta=0.01$ eV is shown in left panel ((a),(c))
			panel while $\Delta=0.1$ eV is depicted in the right ((b),(d)). Two distant power laws
			$s=0.5$ and $s=2$ are shown.}\label{Fig:fig7}
	\end{figure*}
		The microcavity mode induces a `cross-talk' between otherwise disconnected molecules
	(note that we neglected dipole-dipole interactions). This results in a finite
	contribution of phononic displacement in a given molecule, caused by electronically
	excited states residing in others, $\mathcal{X}_{i,e_{j\neq i}}$ (see
	\autoref{Fig:fig6}b), and also a  `molecule-induced' vibrational displacement in the
	cavity state measured by $\mathcal{X}_{i,1}$. Although these contributions are suppressed
	for low frequencies at large $\Omega_R$, in a similar way to the `local' quantity
	$\mathcal{X}_{i,e_i} $, they slightly augment in a counterintuitive way for larger
	frequencies. Non-local and cavity-mode vibrational dressing present very similar trends,
	while the second is less sensitive to the Rabi frequency $\Omega_R$ (see
	\autoref{Fig:fig6}b). Moreover, an analysis varying the ensemble size at large $\Omega_R$
	for $N=1-5$ (not shown) suggests the local vibrational displacement
	$\mathcal{X}_{i,e_{i}}$ scales for the whole frequency spectrum as $1/N$. These results
	generalise the scaling predicted by the variational ansatz solution for the
	HTC~\cite{Zeb2016}, which is related with the prefactor $1/\sqrt{N}$ in the coupling
	between bare electronic states in the LPP. Non-local and cavity-projected vibrational
	dressing also appear to follow a universal trend of $1/N$ in the cases analysed here.
	Therefore, despite the indirect connection of molecular vibrations through the cavity in
	strong coupling, vibrational dressing is collectively suppressed. At moderate Rabi
	frequency, the situation is different, and the reduction with $N$ depends
	non-analytically on $\Omega_R$ and $\Delta$. Although a full analysis is beyond the scope
	of this contribution, we note that the related SBM possesses a set of quantum phase
	transitions for Ohmic and sub-Ohmic baths in which strong polaronic dressing completely
	suppresses tunnelling dynamics. For sub-Ohmic baths, the critical coupling is a function
	of quantities related to $\Omega_R$ and $\Delta$ (see \autoref{sec:single_mol})
	\cite{Leggett1987,Bera2014}, and the ground state contains complex superpositions of
	displaced oscillator states \cite{Bera2014}, which may be relevant for the physics of
	intermediate Rabi couplings.

		As for the bare molecules, this observable is extremely sensitive to the molecular
	species supporting the LPP, even when the Stokes shift is kept constant. Additional insight into
	the strength of the RVD effect for a given 	molecular ensemble is given by the ratio
	\begin{equation}
	\mathcal{R}(\omega_k)=\frac{\mathcal{X}_{i,e_i}(\omega_k,\Omega_R',\Delta)}{\mathcal{X}_{i,e_i}(\omega_k,\Omega_R,\Delta)}, \label{eq:ratio}
	\end{equation}
	where $\Omega_R'>\Omega_R$ are two Rabi frequencies. Intriguingly, as shown in \autoref{Fig:fig7}, the ratio $\mathcal{R}$ is mostly
	independent of the power law in $J_v(\omega)$, unveiling another universal characteristic
	of the model. Therefore, although the LPP is more polaronic for lower $s$, the
	suppression at low frequencies is increased in absolute terms in such a way that the
	\textit{relative suppression} spectrum is molecule-independent.
	
	Exploiting the similarities with the SBM for a single molecule, as discussed in Appendix
	\ref{sec:single_mol}, the intrinsic frequency dependence of $\mathcal{R}$ follows
	from the variational polaron theory for the SBM ($N=1$) with `bias' $\delta=-\Delta$
	(provided that $\omega_e=\omega_O$).  For large $\Omega_R$ and small $\Delta$, bias
	effects caused by the vibrational reorganisation energy become irrelevant. The slow modes
	are anti-adiabatic from the perspective of the Rabi oscillations. Hence, they cannot
	readjust their displacements rapidly enough to track the formation of polaritons arising
	from light-matter coupling~\cite{Chin2011}, and are unable to maximise the
	vibrational dressing of the electronic states. Vibrational and photonic processes are
	hence `decoupled' and $\mathcal{R}$ decreases monotonically with the frequency. This
	picture agrees with the smaller energy shifts for Ohmic and sub-Ohmic spectral densities,
	corresponding to `slow baths'  (see red and blue curves in \autoref{Fig:fig6}a     and
	\autoref{Fig:fig2}a). Conversely, larger exponents $s$ correspond to spectral densities
	where high-frequency modes are dominant. In particular, these modes are adiabatic from
	the $\mathcal{S}$ subsystem perspective and lead to larger frequency shifts and phononic
	displacements. Besides, such effects could induce renormalisation of the coupling
	parameters, which could be observed for instance in a reduced Rabi splitting in linear
	response spectrum.
   	\begin{figure*}[tb]
   	\centering
   	\includegraphics[width=0.95\linewidth]{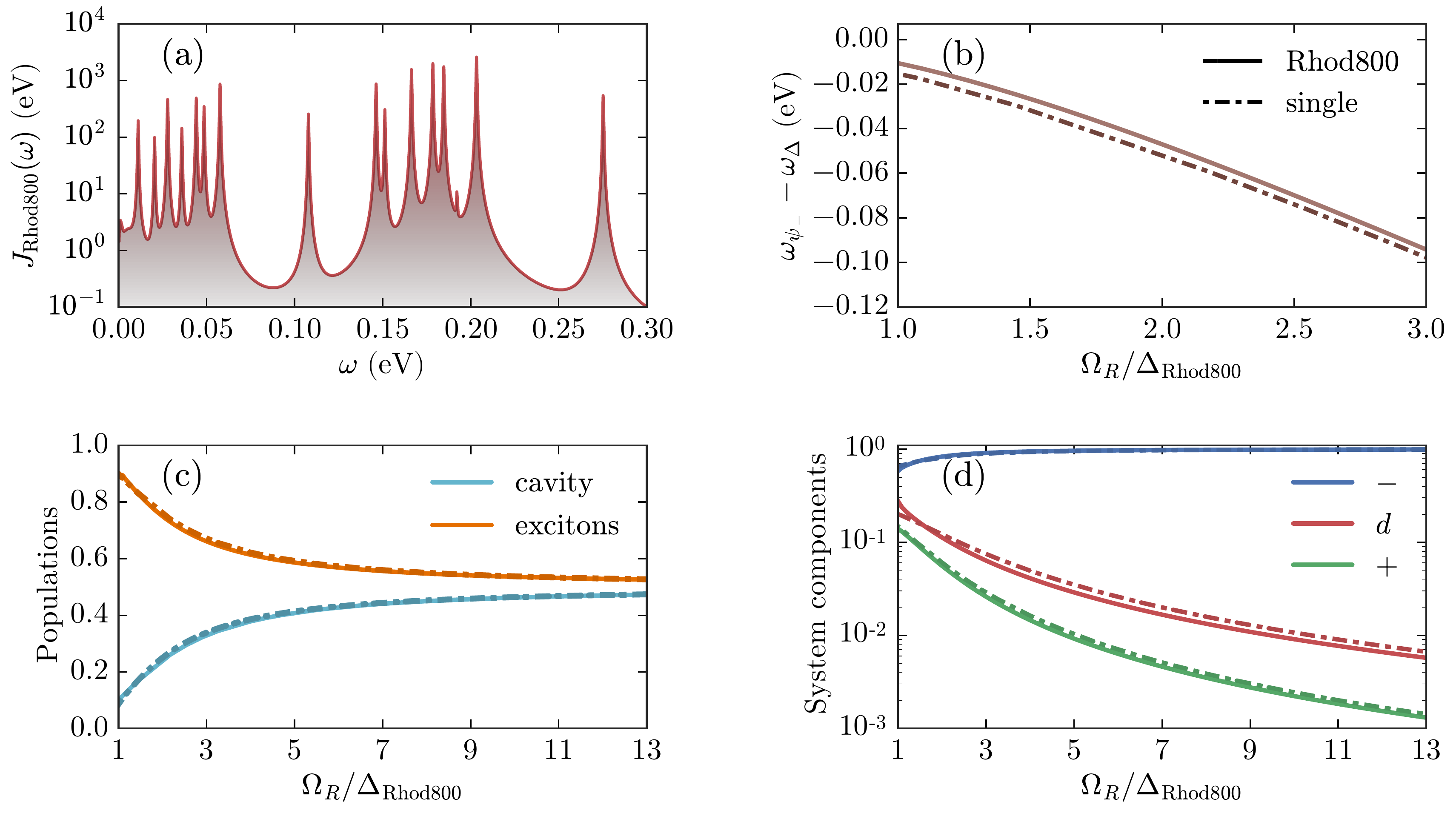}
   	\caption{Reduced observables for the LPP supported by a dimer of Rhod800
   		molecules. The total reorganisation energy for the modes that are considered
   		is $\Delta=0.112$ eV. The numerical first moment of the spectral density
   		$J_{\mathrm{Rhod800}}(\omega)$, depicted in panel (a) reveals a
   		value of $\Omega_v=0.154$ eV.}\label{Fig:fig8}
    \end{figure*}
	The results for $N=5$ molecules displayed in \autoref{Fig:fig7} (lower panels) show a
	more significant suppression of vibrational displacement compared with the SBM. In this
	case, RVD occurs by the additive contribution of the light-matter coupling, described
	above, and ensemble effects such as the `collective protection' due to each molecule in the polariton state staying mostly in its ground state \cite{Galego2016}. Intriguingly, a global minimum in $\mathcal{R}$ is
	observed for a finite $\omega_k\sim\mathcal{O}(\Omega_R)$, suggesting non-trivial bias
	effects due to non-negligible reorganisation energy $\Delta$~\cite{Bera2014}. Indeed, 
	it has recently been shown that the very large displacements associated with the strongly
	coupled slow modes in the sub-Ohmic spin-boson model may develop over very long timescales
	when the system is biased and then prepared out of equilibrium, i.e. by laser excitation~\cite{Eckel2009,Gonzalez-Ballestero2017}.
	This creates an effective time-dependent bias which, as our static results suggest, could dynamically
	alter polaritonic properties in real-time and generate strongly non-Markovian dynamics that might
	be detectable in time-resolved microcavity experiments.  

	The results outlined in this section clearly suggest that the HTC model is insufficient
	to address the non-trivial vibrational features arising for different molecular systems
	($J_v(\omega)$) in the LPP, but that instead it is necessary to take into account the
	vibrational structure of the molecule. Nevertheless, we have shown that the
	relative effect of RVD is somewhat irrespective of the spectral density of molecular
	phonons but is mostly determined by the reaction coordinate frequency and the
	reorganisation energy. 
	
	Up to now, we have considered smooth vibrational spectral densities. In the following, we show how these ideas can be tested in a molecule with highly structured vibrational features that recreates the conditions found
	in realistic organic microcavities.
	\begin{figure*}[tb]
		\centering
		\includegraphics[width=0.95\linewidth]{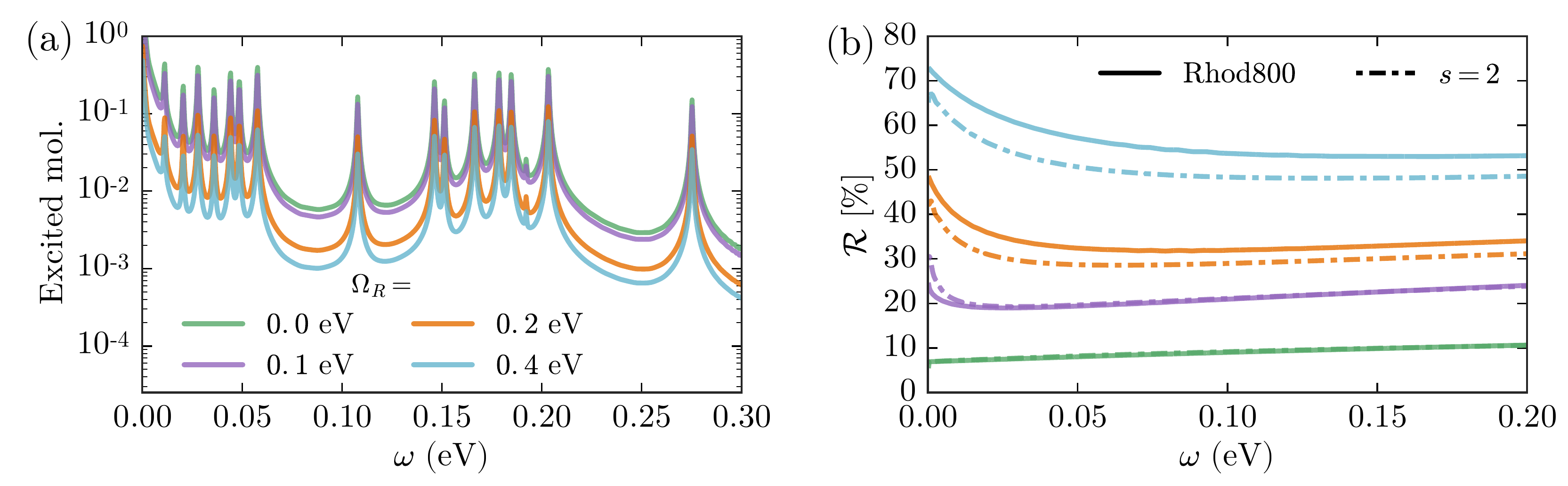}
		\caption{(a) Displacement spectrum for the LPP supported by a dimer of
			Rhod800 molecules, projected on the electronically excited molecule. (b)
			Ratio between the displacement of the electronically excited molecule at
			$\Omega_R'=1.5$ eV and the cases depicted in (a). The remaining parameters
			are fixed according to \autoref{Fig:fig8} caption.}\label{Fig:fig9}
	\end{figure*}
	\section{The Rhodamine 800 molecule}\label{sec:rhod800_GS}
			In typical organic molecules, besides simple power-law spectra at low frequencies, more
	complicated features may arise at high (typically mid-infrared) frequencies, as observed
	in, e.g., infrared and Raman spectroscopy where many sharp vibrational resonances are
	present. In the following, we consider the Rhodamine 800 perchlorate molecular compound
	(sketched in \autoref{Fig:fig8}b), which constitutes a common choice for a laser gain
	media and strong coupling experiments~\cite{Valmorra2011}. In this case, the phononic
	information necessary to `feed' $J_v(\omega)$ is known from vibrational
	spectroscopy~\cite{Christensson2010}. In particular, the data from three pulse photon
	echo peak shift (3PEPS) experiments at room temperature (in Fourier space) can be
	interpolated to give an accurate approximation for the low-frequency part
	$J_v^{\mathrm{lf}}(\omega)$~\cite{Passino1997}. Regarding the high-frequency modes,
	denoted as $\Omega_{k}$, we consider that each oscillator is broadened by an amount of
	$\Gamma\ll\Omega_k$ due to the interaction with a background Ohmic bath of vibrations
	accounting for vibrational dephasing and decay~\cite{Garg1985}, by taking
	\begin{equation}
	J_v^{\mathrm{hf}}(\omega,\Omega_k,\gamma_k)=\pi\frac{4\gamma\omega\Omega_{k}^2\eta^2}{(\Omega_{k}^2-\omega^2)^2+(2\pi\gamma_k\Omega_{k}\omega)^2},
	\end{equation}
	where $\gamma_k=\Gamma/(2\pi\Omega_k)$ is a dimensionless measure of the mode broadening.
	Firstly, we compared the $\mathcal{S}$ populations of a dimer of Rhod800, with spectral
	density $J_{\mathrm{Rhod800}}(\omega)= J_v^{\mathrm{lf}}(\omega) +
	\sum_{k\in\mathrm{hf}}J_v^{\mathrm{hf}}(\omega,\Omega_k,\gamma_k)$, shown in
	\autoref{Fig:fig8}a, with a HTC model with parameters
	$\Omega_v^{\mathrm{Rhod800}},\Delta^{\mathrm{Rhod800}}$ estimated from the experimental
	spectral density. Strikingly, despite the involved resonant structure of
	$J_{\mathrm{Rhod800}}(\omega)$, the energy shift and the reduced density matrix
	observables present a smooth behaviour as the Rabi frequency $\Omega_R$ is varied
	(\autoref{Fig:fig8}), which mimics the HTC model. This predicts that the probing of
	excitonic and photonic characteristics of the LPP supported by an ensemble of arbitrarily
	complex molecules, close to the mechanical equilibrium, only requires the experimental
	knowledge of few collective quantities.

	However, in agreement with the conclusions drawn for the polynomial spectral densities in
	\autoref{eq:Leggett}, the vibrational features cannot be described by the HTC model but
	are profoundly dominated by the vibrational structure of the molecule and the host
	medium, encoded in $J_v(\omega)$. Phononic displacements of the exciton component of the
	LPP in \autoref{Fig:fig9}a, mirror the resonant structures in    
	$J_{\mathrm{Rhod800}}(\omega)$. Nevertheless, the relative suppression $\mathcal{R}$ is
	surprisingly smooth see \autoref{Fig:fig9}b. In agreement with our previous results, is controlled by just the reaction coordinate frequency
	$\Omega_v$ and the reorganisation energy $\Delta$ instead of details in $J_v(\omega)$.

	\section{Conclusion}    
		In this work, we have investigated a microscopic theory that goes beyond current models
	for the lower polaron polariton in organic microcavities, by introducing the whole
	phononic spectrum. In order to treat the problem numerically, we employed a quasi-exact
	VMPS algorithm, able to handle the full Hamiltonian including multimode and many molecule
	effects. We have demonstrated that the RVD of polaritons compared to bare electronic
	states is a universal feature of
	strongly-coupled organic microcavities that does not depend on the details of the vibrational structure.
	In this way, different distributions of intra- or extra- molecular vibrational modes lead
	to similar excitonic and photonic properties of the LPP wavefunction, where only the bath
	component is appreciably altered. 
	
		Our findings show unambiguosly how, despite the involved phononic spectrum in organic
	molecules, a large extent of features in the LPP can still be emulated by the HTC model.
	The universal character of the results is not expected to be present in the time
	evolution of the exciton-photonic system, a problem that has attracted interest lately
	due to the experimental advances achieving exotic organic polariton
	dynamics~\cite{Schwartz2013} and non-equilibrium BEC in the strong coupling
	regime~\cite{Rodriguez2013}. As we show in~\cite{DelPino2018} the non-equilibrium
	description of the bath modes is crucial to determine the time evolution route followed
	by the system arising from details in $J_v(\omega)$~\cite{Chin2013}.
	
	However, there is still a long way to go regarding experiments to
	obtain conclusive evidence of changes in chemical processes or electronic
	energy transfer in ESC. The model presented and explored here seeks to generalise the
	most simple descriptions of the problem, while future prospects along these lines would
	go towards extended approaches, that are able to introduce multi-dimensional features in
	the recent field of polaritonic chemistry.

	\section*{Acknowledgments}
	This work has been funded by the European Research Council (ERC-2011-AdG-290981
	and ERC-2016-STG-714870), and the Spanish MINECO under
	contract MAT2014-53432-C5-5-R and the ``María de Maeztu'' programme for Units of
	Excellence in R\&D (MDM-2014-0377). F.A.Y.N.S. and A.W.C. gratefully acknowledge
	the support of the Winton Programme for the Physics of Sustainability and
	EPSRC\@.
	
	\bibliography{references}
	
	\appendix
	
    \section{Tensor-Network-based approach}
	
	In the following appendix, we outline the numerical approach to target the optimal LPP
	$\ket{\psi_-}$. We state only what is necessary to understand this work, referring to the
	reader to complete reviews to the topic~\cite{Orus2014, Schollwock2011}.
	
	The numerical approach is based on the two following ideas: \emph{(i)} Representation of
	$\ket{\psi_-}$ as a multi-chain TN, mimicking the structure in the star Hamiltonian
	\autoref{Fig:fig1} that can be employed to implement the \emph{(ii)} variational
	principle
	\begin{equation}
	\ket{\psi_-}=\min_{\ket{\psi}\in\mathcal{M}_{\mathrm{TN}}}\frac{\braket{\psi|\hat{H}_\mathcal{S}+\hat{H}_v|\psi}}{\braket{\psi|\psi}},\label{eq:var_opt}
	\end{equation}
	within the TN sub-manifold of the total
	Hilbert space ($\mathcal{M}_{\mathrm{TN}}\subset\mathcal{H}_{\mathcal{S}}\otimes\mathcal{H}_{\mathcal{E}_v}$) that leads to the star-DMRG algorithm employed in the work.
	
	\subsection{Star TN structure}\label{sec:TN_approach}
	The DMRG algorithm works especially well targeting ground states in one-dimensional systems. The orthogonal
	mapping of the vibrational modes, leading to the star Hamiltonian introduced in the main
	text, constitutes a convenient starting point for the application of this numerical
	method. We consider the expansion of the LPP in the single exciton-photon subspace, which
	exploits the Fock states $\ket{n_l^{(i)}}\in\mathcal{E}_v^{(i)}$ for the $l$-th site at
	chain $i$, i.e.
	\begin{equation}
	\ket{\psi_-} =\sum_{n_\mathcal{S}=1}^{N+1}\sum_{i=1}^{N}\sum_{\boldsymbol{n}^{(i)}}\Psi_{n_\mathcal{S},\boldsymbol{n}^{(i)}}\ket{n_\mathcal{S}}\otimes\ket{\boldsymbol{n}^{(i)}},
	\end{equation}
	where the array $\boldsymbol{n}^{(i)}=(n_0^{(i)},\ldots,n_{L-2}^{(i)})$ encompasses the Fock numbers
	at each of the chains. This state admits a representation as an TN\@. In this case the complex
	$M_v+1$-th order tensor $\Psi_{n_\mathcal{S},\boldsymbol{n}^{(1)}, \ldots, \boldsymbol{n}^{(N)}}$ is
	decomposed into a product of rectangular matrices with site-dependent dimensions $d_l$
	under the following setting: each of the excitations of a given bosonic site and chain
	$n_l^{(i)} \in \big(0, 1, \ldots, \infty\big)$ is assigned to a matrix
	$A_{d_{l-1},d_l}^{n_l^{(i)}} \in \mathbb{C}^{D_{l-1}\times D_l}$, while each of the
	system degrees of freedom corresponds to $A_{1,d_1}^{n_\mathcal{S}}\in\mathbb{C}^{1\times
		D_\mathcal{S}}$. Inherited from the star Hamiltonian depicted \autoref{Fig:fig1}, the
	most natural TN representation of $\ket{\psi_-}$ that enables to reduce the  minimisation  \autoref{eq:var_opt}
	to a single site problem, in the way described in \autoref{sec:DMRG_alg} is given by
	\begin{equation}
	\ket{\psi_{-}}=\sum_{n_{\mathcal{S}}=1}^{N+1}\textbf{A}^{n_{\mathcal{S}}}\sum_{i=1}^{N}\sum_{\boldsymbol{n}_i}\big(\textbf{A}^{n^{(i)}_{0}}\ldots\textbf{A}^{n^{(i)}_{M_v-1}}\big)\ket{n_\mathcal{S}}\otimes\ket{\boldsymbol{n}^{(i)}}.
	\label{eq:MPS}
	\end{equation}
	The order in which the \textbf{A} matrices occur in the product, sketched in
	\autoref{Fig:A1}a, mimics how the sites are connected in the Hamiltonian
	\autoref{eq:H_chain}. Instead, here the system tensors (site 0) corresponding to the
	$0^{\mathrm{th}}$ site are contracted with the boson tensor at each of the
	$1^{\mathrm{st}}$ sites. To account for bosonic degrees of freedom, a decomposition of
	the bath modes in terms of an optimal boson basis (OBB) is
	considered~\cite{Guo2012,Schroder2016}. The auxiliary indexes that are contracted via the
	matrix product, $d_l \in (1, 2, \ldots, D_l)$ are known as bond dimensions. They embody
	the entanglement content of the state; higher bond dimensions signify more
	entanglement~\cite{Orus2014}, such that for sufficiently high $D_l$ the form
	\autoref{eq:MPS} can represent any quantum state in the Hilbert space.

The major power of TN-based approaches relies on the efficient computer representation of
the site tensors $\textbf{A}$. In practice, this `compression' is achieved by
successive iterations of singular value decompositions and truncation of the states with
small singular values, corresponding to the Schmidt coefficients~\cite{Schollwock2011}.
This protocol reduces the maximum bond dimensions, enabling to restrict the possible
states to a sub-manifold $\mathcal{M}_{\mathrm{TN}} \subset
\mathcal{H}_\mathcal{S}\otimes\mathcal{H}_{\mathcal{E}_v}$ that comprises the low-lying
many-body states of the system.

\subsection{Star DMRG algorithm}\label{sec:DMRG_alg}

\begin{figure*}[tb]
	\includegraphics[width=\linewidth]{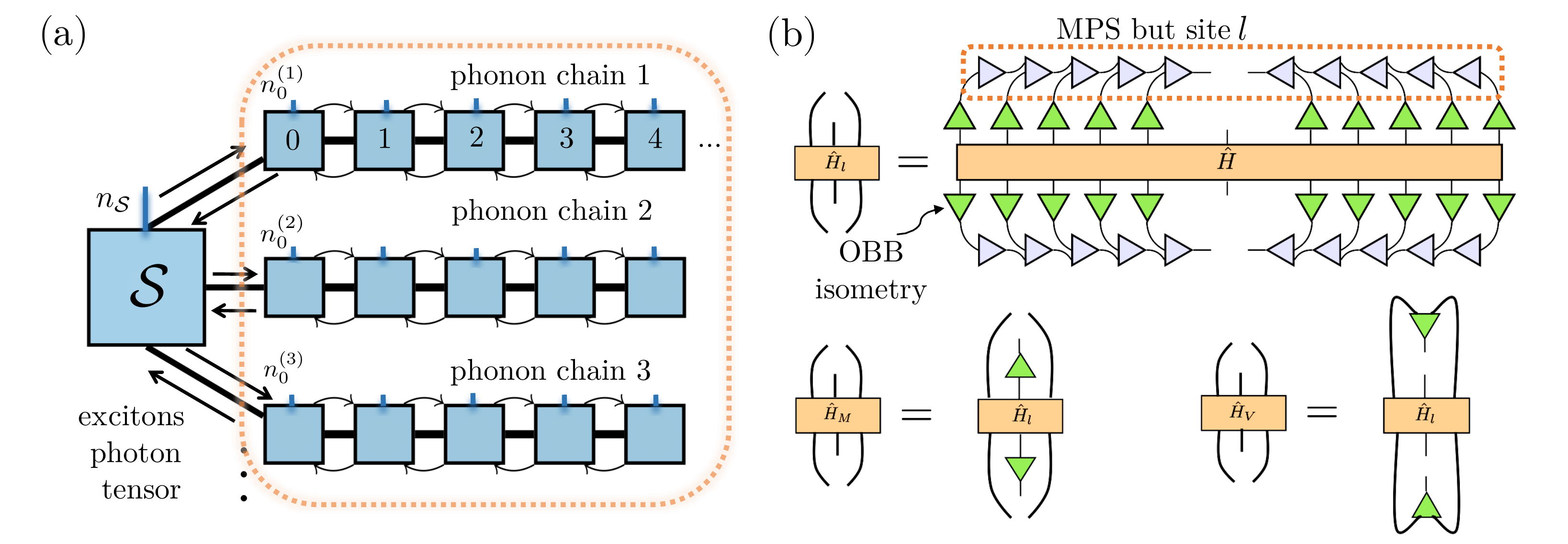}
	\caption{(a) Sketch of the star TN for the case $N=3,M_v=5$ that
		is used to represent the LPP\@. Here each of the squares signifies a tensor,
		that is contracted with the tensor to which its legs are connected. Arrows depict a possible order at which the tensors are optimised in the algorithm.
		The open legs represent the local Hilbert space at each of the sites, which is
		contracted with the Hamiltonian $\hat{H}=\hat{H}_\mathcal{S}+\hat{H}_v$ to state the minimisation problem.
		To carry out this minimisation, it is instrumental to represent the state in \emph{mixed canonical form} compute the contractions depicted in (b). Here we use a similar notation for the tensors as in the
		reference~\cite{Haegeman2011}.}\label{Fig:A1}
\end{figure*}

Employing the star TN representation \autoref{eq:MPS} in conjunction with the star
Hamiltonian, the minimisation problem \autoref{eq:var_opt}, limited to the manifold
$\mathcal{M}_{\mathrm{TN}}$ can be performed by a DMRG-sweeping procedure, which
optimises one matrix at a time while keeping all others fixed, then optimising the
neighbouring matrix, and so forth, until convergence is achieved. The basic steps of the
algorithm employed, based on sequential 1D DRMG~\cite{Schollwock2011} sweeps along the
chain, are sketched in \autoref{Fig:A1}a.

To allow for fast contractions we keep the state in a mixed canonical form, enabling to
benefit from orthogonality conditions as shown in~\cite{haegeman2016unifying}. Such
representation reduces each DMRG step as a linear eigenvalue (optimisation) problem for a
single site tensor (either $\textbf{A}_l$ or the OBB matrices) with the effective
Hamiltonians shown in \autoref{Fig:A1}b. More details about the
numerical implementation can be found in~\cite{Schroder2016}.

Finally, regarding the convergence of the DMRG algorithm for dense spectra where
$M_v\gg1$, is important to notice that for gap-less finite-bandwidth $J_v(\omega)$, it
has been shown~\cite{Chin2010}, the chain becomes asymptotically homogeneous, such that
$\omega_{l\gg1}\rightarrow\omega_c/2$ and $t_{l\gg1}\rightarrow\omega_c/4$ respectively.
The translational-invariant slice of the chain can be diagonalised in `chain momentum'
space, giving rise to a cosine energy dispersion band $\sim\omega_c(1-\cos\pi q)$ with a
minimum at zero energy. Thus there is no net contribution of these sites to the LPP
energy, and the truncated chain approximation considered in this work is fulfilled.

\section{Single-molecule multimode limit}\label{sec:single_mol}

One interesting case of the Hamiltonian under study is the single-molecule limit. In
particular, single-molecule strong coupling has been recently reported in plasmonic
nanocavities~\cite{Zengin2013,Chikkaraddy2016}. Within the single electronic/photonic
excitation subspace, this case can be mapped exactly to the well-known Spin Boson Model
(SBM)~\cite{Leggett1987, Weiss1999} by a shift of the vibrational mode origin in the
original Hamiltonian \autoref{eq:Our_H} before performing the chain transformation. After
this shift, described by $\hat{H}_v^{\mathrm{shift}}= e^{\hat{C}}\hat{H}_ve^{-\hat{C}}$
with $\hat{C} = \sum_k \lambda_k (\hat{b}_k-\hat{b}_k^{\dagger})/(2\omega_k)$, the
light-matter coupling can be expressed through the dynamics of a quasispin
$\hat{\Sigma}_-=\ket{e}\bra{1}$ coupled to a bath of bosons, governed by the Hamiltonian
$\hat{H}_{\mathrm{SBM}}=\hat{H}_\mathcal{S}+\hat{H}_v^{\mathrm{shift}}$
\begin{equation}\label{eq:SBM_eff}
\hat{H}_{\mathrm{SBM}}=\frac{\delta}{2}\hat{\Sigma}_{z}+\frac{g}{2}\hat{\Sigma}_{x}+\sum_k\big[\omega_{k}\hat{n}_k + \frac{\lambda_k}{2} \hat{\Sigma}_{z} (\hat{b}_k+\hat{b}_k^{\dagger})\big].
\end{equation}
Here, we have have introduced $\ket{e}=\hat{\sigma}^+\ket{G}$ and $\ket{1} = \hat{a}^\dag
\ket{G}$ as shortcuts for the excitonic and photonic states in the single-excitation
subspace, respectively.

The SBM constitutes one of the minimal models to study quantum dissipation in solid-state
and organic systems, e.g. decoherence of quantum oscillations in
qubits~\cite{Costi2003, Khveshchenko2004}, impurity moments coupled to bulk magnetic
fluctuations~\cite{Sachdev1999}, and electron transfer in biological
molecules~\cite{Plenio2008}. Here, the detuning between the cavity and the zero-phonon
line $\delta=\omega_{e}-\Delta-\omega_{O}$ plays the role of a bias, where the
reorganisation energy follows from \autoref{eq:reorg}.

The light-matter coupling $g$ favours the mixing of the molecular exciton and the photon,
similarly to the HTC model analysed in previous section, whereas the spin-bath
interaction dresses them with vibrational modes, destroying light-matter coherence. In
the unbiased case, i.e., when the cavity is resonant with the zero-phonon line,
$\omega_O=\omega_e-\Delta$, the Hamiltonian becomes parity symmetric under the exchange
$\ket{e} \Leftrightarrow \ket{1}$. This symmetry is known to be spontaneously broken when
the vibrational coupling becomes larger than a critical value for sub-Ohmic and Ohmic
$J_v(\omega)$, leading to a quantum phase transition in which the ground state is a fully
polarised spin state~\cite{Sachdev2007}. Translated to the present case, this would
signify a novel vibrationally-driven localisation phenomenon in either excitonic or
photonic states, precluding the formation of polaritons at the single molecule limit.
However, it is questionable whether traces of this transition could be observed in
nanocavity systems capable of significant light-matter coupling in the single-molecule
case~\cite{Chikkaraddy2016, Liu2017}, especially considering that we are here neglecting
the nonradiative and radiative losses associated with such systems. We thus have focused on 
the many-molecule case ($N>1$) at zero detuning $\omega_e=\omega_O$,
corresponding to the most common experimental setups, while using the analogy to the SBM
to make connections to the existing literature.
	
\end{document}